\documentclass[aps,prc,superscriptaddress,showpacs,nofootinbib,floatfix,twocolumn,amsmath,amssymb]{revtex4}
\usepackage{graphicx}

\newcommand{\sqrts}{\mbox{$\sqrt{s_{_{NN}}}$}}
\newcommand{\pt}{\mbox{$p_{\rm T}$}}
\newcommand{\gevc}{\mbox{GeV/$c$}}

\newcommand{\auau}{\mbox{Au$+$Au}}
\newcommand{\pp}{\mbox{$p+p$}}
\newcommand{\dau}{\mbox{$d+$Au}}
\newcommand{\pizero}{\mbox{$\pi^0$}}

\begin{document}

\title{Centrality dependence of charged hadron production \\
	in deuteron+gold and nucleon+gold collisions at $\sqrt{s_{NN}}$ = 200 GeV} 

\newcommand{\abilene}{Abilene Christian University, Abilene, TX 79699, USA}
\newcommand{\acadsin}{Institute of Physics, Academia Sinica, Taipei 11529, Taiwan}
\newcommand{\banaras}{Department of Physics, Banaras Hindu University, Varanasi 221005, India}
\newcommand{\barc}{Bhabha Atomic Research Centre, Bombay 400 085, India}
\newcommand{\bnl}{Brookhaven National Laboratory, Upton, NY 11973-5000, USA}
\newcommand{\caucr}{University of California - Riverside, Riverside, CA 92521, USA}
\newcommand{\ciae}{China Institute of Atomic Energy (CIAE), Beijing, People's Republic of China}
\newcommand{\cns}{Center for Nuclear Study, Graduate School of Science, University of Tokyo, 7-3-1 Hongo, Bunkyo, Tokyo 113-0033, Japan}
\newcommand{\colorado}{University of Colorado, Boulder, CO 80309, USA}
\newcommand{\columbia}{Columbia University, New York, NY 10027 and Nevis Laboratories, Irvington, NY 10533, USA}
\newcommand{\dapnia}{Dapnia, CEA Saclay, F-91191, Gif-sur-Yvette, France}
\newcommand{\debrecen}{Debrecen University, H-4010 Debrecen, Egyetem t{\'e}r 1, Hungary}
\newcommand{\elte}{ELTE, E{\"o}tv{\"o}s Lor{\'a}nd University, H - 1117 Budapest, P{\'a}zm{\'a}ny P. s. 1/A, Hungary}
\newcommand{\fsu}{Florida State University, Tallahassee, FL 32306, USA}
\newcommand{\gsu}{Georgia State University, Atlanta, GA 30303, USA}
\newcommand{\hiroshima}{Hiroshima University, Kagamiyama, Higashi-Hiroshima 739-8526, Japan}
\newcommand{\ihepprot}{IHEP Protvino, State Research Center of Russian Federation, Institute for High Energy Physics, Protvino, 142281, Russia}
\newcommand{\illuiuc}{University of Illinois at Urbana-Champaign, Urbana, IL 61801, USA}
\newcommand{\isu}{Iowa State University, Ames, IA 50011, USA}
\newcommand{\jinrdubna}{Joint Institute for Nuclear Research, 141980 Dubna, Moscow Region, Russia}
\newcommand{\kek}{KEK, High Energy Accelerator Research Organization, Tsukuba, Ibaraki 305-0801, Japan}
\newcommand{\kfki}{KFKI Research Institute for Particle and Nuclear Physics of the Hungarian Academy of Sciences (MTA KFKI RMKI), H-1525 Budapest 114, POBox 49, Budapest, Hungary}
\newcommand{\korea}{Korea University, Seoul, 136-701, Korea}
\newcommand{\kurchatov}{Russian Research Center ``Kurchatov Institute", Moscow, Russia}
\newcommand{\kyoto}{Kyoto University, Kyoto 606-8502, Japan}
\newcommand{\labllr}{Laboratoire Leprince-Ringuet, Ecole Polytechnique, CNRS-IN2P3, Route de Saclay, F-91128, Palaiseau, France}
\newcommand{\lawllnl}{Lawrence Livermore National Laboratory, Livermore, CA 94550, USA}
\newcommand{\losalamos}{Los Alamos National Laboratory, Los Alamos, NM 87545, USA}
\newcommand{\lpc}{LPC, Universit{\'e} Blaise Pascal, CNRS-IN2P3, Clermont-Fd, 63177 Aubiere Cedex, France}
\newcommand{\lund}{Department of Physics, Lund University, Box 118, SE-221 00 Lund, Sweden}
\newcommand{\muenster}{Institut f\"ur Kernphysik, University of Muenster, D-48149 Muenster, Germany}
\newcommand{\myongji}{Myongji University, Yongin, Kyonggido 449-728, Korea}
\newcommand{\nagasaki}{Nagasaki Institute of Applied Science, Nagasaki-shi, Nagasaki 851-0193, Japan}
\newcommand{\newmex}{University of New Mexico, Albuquerque, NM 87131, USA }
\newcommand{\nmsu}{New Mexico State University, Las Cruces, NM 88003, USA}
\newcommand{\ornl}{Oak Ridge National Laboratory, Oak Ridge, TN 37831, USA}
\newcommand{\orsay}{IPN-Orsay, Universite Paris Sud, CNRS-IN2P3, BP1, F-91406, Orsay, France}
\newcommand{\peking}{Peking University, Beijing, People's Republic of China}
\newcommand{\pnpi}{PNPI, Petersburg Nuclear Physics Institute, Gatchina, Leningrad region, 188300, Russia}
\newcommand{\riken}{RIKEN (The Institute of Physical and Chemical Research), Wako, Saitama 351-0198, JAPAN}
\newcommand{\rikjrbrc}{RIKEN BNL Research Center, Brookhaven National Laboratory, Upton, NY 11973-5000, USA}
\newcommand{\saopaulo}{Universidade de S{~a}o Paulo, Instituto de F\'{\i}sica, Caixa Postal 66318, S{~a}o Paulo CEP05315-970, Brazil}
\newcommand{\seoulnat}{System Electronics Laboratory, Seoul National University, Seoul, South Korea}
\newcommand{\stonybrkc}{Chemistry Department, Stony Brook University, Stony Brook, SUNY, NY 11794-3400, USA}
\newcommand{\stonycrkp}{Department of Physics and Astronomy, Stony Brook University, SUNY, Stony Brook, NY 11794, USA}
\newcommand{\subatech}{SUBATECH (Ecole des Mines de Nantes, CNRS-IN2P3, Universit{\'e} de Nantes) BP 20722 - 44307, Nantes, France}
\newcommand{\tenn}{University of Tennessee, Knoxville, TN 37996, USA}
\newcommand{\titech}{Department of Physics, Tokyo Institute of Technology, Oh-okayama, Meguro, Tokyo 152-8551, Japan}
\newcommand{\tsukuba}{Institute of Physics, University of Tsukuba, Tsukuba, Ibaraki 305, Japan}
\newcommand{\vandy}{Vanderbilt University, Nashville, TN 37235, USA}
\newcommand{\waseda}{Waseda University, Advanced Research Institute for Science and Engineering, 17 Kikui-cho, Shinjuku-ku, Tokyo 162-0044, Japan}
\newcommand{\weizmann}{Weizmann Institute, Rehovot 76100, Israel}
\newcommand{\yonsei}{Yonsei University, IPAP, Seoul 120-749, Korea}
\affiliation{\abilene}
\affiliation{\acadsin}
\affiliation{\banaras}
\affiliation{\barc}
\affiliation{\bnl}
\affiliation{\caucr}
\affiliation{\ciae}
\affiliation{\cns}
\affiliation{\colorado}
\affiliation{\columbia}
\affiliation{\dapnia}
\affiliation{\debrecen}
\affiliation{\elte}
\affiliation{\fsu}
\affiliation{\gsu}
\affiliation{\hiroshima}
\affiliation{\ihepprot}
\affiliation{\illuiuc}
\affiliation{\isu}
\affiliation{\jinrdubna}
\affiliation{\kek}
\affiliation{\kfki}
\affiliation{\korea}
\affiliation{\kurchatov}
\affiliation{\kyoto}
\affiliation{\labllr}
\affiliation{\lawllnl}
\affiliation{\losalamos}
\affiliation{\lpc}
\affiliation{\lund}
\affiliation{\muenster}
\affiliation{\myongji}
\affiliation{\nagasaki}
\affiliation{\newmex}
\affiliation{\nmsu}
\affiliation{\ornl}
\affiliation{\orsay}
\affiliation{\peking}
\affiliation{\pnpi}
\affiliation{\riken}
\affiliation{\rikjrbrc}
\affiliation{\saopaulo}
\affiliation{\seoulnat}
\affiliation{\stonybrkc}
\affiliation{\stonycrkp}
\affiliation{\subatech}
\affiliation{\tenn}
\affiliation{\titech}
\affiliation{\tsukuba}
\affiliation{\vandy}
\affiliation{\waseda}
\affiliation{\weizmann}
\affiliation{\yonsei}
\author{S.S.~Adler}	\affiliation{\bnl}
\author{S.~Afanasiev}	\affiliation{\jinrdubna}
\author{C.~Aidala}	\affiliation{\columbia}
\author{N.N.~Ajitanand}	\affiliation{\stonybrkc}
\author{Y.~Akiba}	\affiliation{\kek} \affiliation{\riken}
\author{A.~Al-Jamel}	\affiliation{\nmsu}
\author{J.~Alexander}	\affiliation{\stonybrkc}
\author{K.~Aoki}	\affiliation{\kyoto}
\author{L.~Aphecetche}	\affiliation{\subatech}
\author{R.~Armendariz}	\affiliation{\nmsu}
% ...
\collaboration{PHENIX Collaboration} \noaffiliation

\date{\today}

\begin{abstract}

We present transverse momentum ($p_{T}$) spectra of charged 
hadrons measured in deuteron-gold and nucleon-gold collisions at 
\sqrts\ = 200 GeV for four centrality classes.  Nucleon-gold 
collisions were selected by tagging events in which a spectator 
nucleon was observed in one of two forward rapidity detectors. The 
spectra and yields were investigated as a function of the number of 
binary nucleon-nucleon collisions, $\nu$, suffered by deuteron 
nucleons.  A comparison of charged particle yields to those in \textit{p}+\textit{p} 
collisions show that yield per nucleon-nucleon collision saturates 
with $\nu$ for high momentum particles. We also present the charged 
hadron to neutral pion ratios as a function of $p_{T}$.

\end{abstract}


\maketitle

\section{Introduction\label{sec:intro}}

The measurement of hadron spectra at high transverse momenta ($p_{T}$) has 
been a powerful tool in studying the modification of particle production and 
propagation in the nuclear medium at the BNL Relativistic Heavy Ion Collider (RHIC).  
One of the most intriguing results observed so far at RHIC is the strong 
suppression of the yield of hadrons with \pt\ above $\sim2$ \gevc\ in midcentral 
and central \auau\ collisions relative to the corresponding yield in 
\pp\ collisions scaled by the sum of independent nucleon-nucleon collisions in the \auau\ 
interaction~\cite{ppg003,ppg014,ppg023}. This is consistent with a picture in 
which the hard scattered parton loses significant energy when it traverses the 
medium created during the collision~\cite{en_loss}. There is additional evidence 
for the strong interaction of partons with the final state medium. For example, two 
particle azimuthal correlations demonstrate significant modifications of jets 
such as width and particle content~\cite{jet_modif,ppg033}. In the \pt\ region, 2 
$<$ \pt\ $<$ 5\,GeV/$c$, an anomalously large baryon/meson ratio is 
observed~\cite{Adler:2003kg}. The coalescence of thermal and shower partons 
offers a possible explanation of this phenomenon as a final state 
effect~\cite{recomb}.

The hadron spectrum is sensitive to final state effects including jet quenching, 
parton recombination and scattering of produced particles, in addition to initial 
state effects, such as gluon saturation and nuclear shadowing~\cite{colorglass}, 
and the Cronin effect~\cite{cronin,lowenergy}.  If shadowing or gluon saturation 
are actually responsible for the large suppression seen in \auau\ collisions, 
these same effects would also produce a suppression of 20-30\% in the yield of 
high \pt~\cite{colorglass} hadrons even in \dau\ collisions.

To study the role of initial state effects in the high \pt\ suppression, 
\dau\ collisions at RHIC serve as a control experiment, since no dense matter is 
expected to be formed in these collisions.  Earlier measurements on inclusive 
hadron production in \dau\ collisions show no suppression of high 
\pt\ hadrons~\cite{minbias,ppg044}.  Rather, there is a small enhancement in the yield 
of high momentum particles in \dau\ collisions.  This result implies that the 
effects seen in central \auau\ collisions are largely due to final state 
interactions in the dense medium.

In this paper, we extend our investigation of charged hadron production in 
\dau\ collisions at \sqrts\ = 200\, GeV at midrapidity with detailed centrality 
selections and with significantly higher statistics. The centrality selection 
allows one to make quantitative statements about the impact parameter dependence 
of particle production.

The nuclear modification of high \pt\ hadron production in proton reactions with 
heavy nuclei has previously been studied at lower center of mass 
energies~\cite{lowenergy, cronin}. It was found that hadron yield at high \pt\ in 
$p$+$A$ collisions increases faster than the nuclear mass, $A$. This effect, 
known as the Cronin effect, is conventionally parametrized as $A^{\alpha}$. At 
$\sqrt{s}\sim$10-30 GeV, $\alpha$ depends on \pt\ and is found to be greater than 
unity for \pt\ $>$ 2\,GeV/$c$.  The exponent $\alpha$ also depends on the particle 
species; it was found to be larger for baryons than for mesons.  The Cronin 
effect is often attributed to the multiple scattering of projectile partons 
propagating through the target nucleus~\cite{lev}.

A deuteron projectile has an additional interesting characteristic: since the average 
separation between the proton and neutron in the deuteron is large, on the order 
of few a fm, there is a class of events in which only one of the two nucleons 
participates in the collision.  Thus \dau\ events can be categorized into 
$``p"$$+$Au or $``n"$$+$Au collisions, when only one of the constituent nucleons 
interacts, and $pn$$+$Au, when both nucleons participate in the collision. In 
such events the interacting proton or neutron is not a free proton or neutron, so 
it is not exactly the same as a $p+$Au or $n+$Au interaction. For simplicity we 
will refer to the tagged samples as $p+$Au and $n+$Au and collectively $N+$Au. 
The selection of $p+$Au and $n+$Au collisions in the PHENIX experiment is done by 
tagging events where the incoming deuteron has a non-interacting neutron or 
proton. Combined with centrality selection, such event tagging provides better 
control of the collision geometry and of the number of subsequent inelastic 
nucleon-nucleon interactions in the gold nucleus.

Very little experimental information has been published about neutron-nucleus 
collisions at high energy. Tagging allows the direct comparison of $pn+$Au, 
$p+$Au, and $n+$Au interactions. Results from these collisions are compared with 
\pp\ data measured by the PHENIX experiment at the same beam energy to shed light on the 
Cronin enhancement and other nuclear effects. Further, we have investigated the 
Cronin effect as a function of the number of collisions per participant nucleon 
in the deuteron. To estimate the fraction of pions in the charged hadron 
spectrum, we have looked at hadron to pion ratios as a function of transverse 
momentum.

The paper is organized as follows.  Section II gives a detailed account of the 
analysis in which we describe the detector, centrality selection, charged particle 
background rejection, spectra corrections and systematic errors. In Sec. III, 
the centrality and \pt\ dependence of the charged hadron spectra, and a comparison with the \pp\ and the \pizero\ data are discussed. A summary is given in Sec. IV.

\section{Experiment and Analysis\label{sec:exp}}
\subsection{Detector\label{sec:detector}} 

The PHENIX detector consists of four spectrometer arms positioned around the 
vertex of colliding deuteron and gold nuclei---two central arms at midrapidity and two muon arms at forward rapidities---and a set of detectors. A detailed 
description of the detector can be found elsewhere, see Ref. ~\cite{NIM} and references 
therein.  This analysis uses the two central arms and the set of global detectors.

Each central arm covers the pseudorapidity range $|\eta|<0.35$, and $90^{\circ}$ in 
azimuth. Charged particles are tracked by drift chambers (DCs) located 2\,m from 
the vertex and layers of pad chambers (PC1 and PC3) positioned at 2.5 and 5\,m 
in the radial direction. The central spectrometer provides axial magnetic fields 
along the beam pipe.  The transverse momentum, \pt\ of each particle is 
determined by its deflection angle in the azimuthal direction as measured by the 
DC. The total particle momentum is reconstructed by projecting tracks back to the 
collision vertex through the magnetic field.  The track reconstruction efficiency 
is approximately 98\% and independent of \pt\ with negligible centrality 
dependence.  The particle momenta are measured with a resolution $\delta p/p = 
0.007\oplus 0.011p$, with $p$ in GeV/$c$. The absolute momentum scale is calibrated to 
0.7$\%$ from the reconstructed proton mass using the PHENIX time-of-flight 
system.  At high transverse momentum, a substantial background of 
electrons is produced by photon conversion in the material between the beam pipe and 
drift chambers. To subtract this background from photon conversion we use the 
ring imaging Cherenkov detector (RICH). The RICH is filled with $\rm CO_2$ gas 
at atmospheric pressure and has a charged particle threshold $\gamma_{\rm{th}}=35$ for 
emission of Cherenkov photons.

\begin{figure}[ht]
\includegraphics[width=1.0\linewidth]{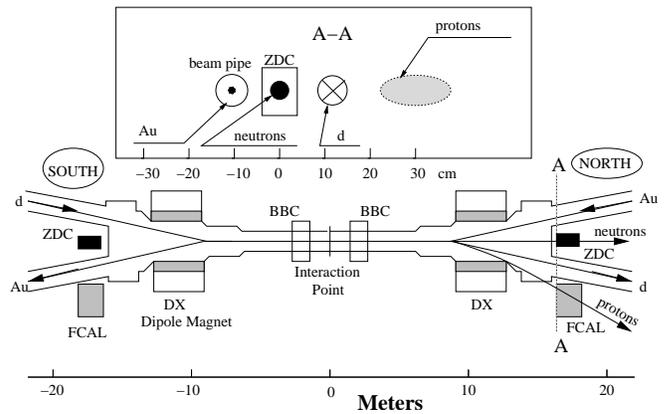}
\caption{\label{fig:zdc_fcal}FCAL, ZDC, and BBC positions relative to the vertex. 
The vertical scale in the figure is arbitrary. 
The insert at the top shows the position of primary beam, spectator neutron
and proton spots at the FCAL and ZDC locations; deuteron beam, spectator 
neutrons and protons go into the plane, the gold beam is coming out of the plane.}
\end{figure}

To characterize the global parameters of the collision and its centrality, the 
PHENIX experiment uses beam-beam counters (BBCs) covering the pseudorapidity range 
3.0$< |\eta| <$ 3.9, zero degree calorimeters (ZDCs) and forward calorimeters 
(FCALs) located at $|\eta|> 6$. The locations of these global detectors are 
schematically shown in Fig.~\ref{fig:zdc_fcal}. Each BBC is an array of 64 
Cherenkov counters around the beam pipe and is positioned at 1.44\,m upstream or 
downstream of the nominal vertex location.  The information from the BBC is used 
for the event timing, vertex position, and centrality determination. The two ZDCs 
are hadronic calorimeters which measure spectator neutrons~\cite{zdc}. They are 
located 18\,m from the interaction point. At the top RHIC energy of 100 
GeV/nucleon, neutrons evaporated from the spectator remnants of the collision are 
emitted within 1 mrad from the colliding beam direction. Charged fragments and 
the noninteracted primary beam are bent by deflecting (DX) magnets to much 
larger angles. The ZDC measures the total neutron energy within a small cone and 
with this provides the number of spectator neutrons from the interacting nucleus. 
The ZDC on the north side measured 100 GeV neutrons from deuteron fragmentation 
with a resolution of $\sigma=$ 28 GeV.
 
The forward calorimeters were installed before the \dau\ run. The FCAL is a hadron 
calorimeter and consists of lead scintillating fiber modules originally used in 
BNL Alternating Gradient Synchotron (AGS) Experiment E864~\cite{e864} rearranged into two 9 by 10 arrays. 
The only difference from the E864 experiment is the readout electronics, which 
are identical to PHENIX central arm electromagnetic calorimeter electronics. The 
size of each module is 10 $\times$ 10 $\times$ 117\,cm, the average tower 
density is 9.6 g/cm$^3$, and the total length corresponds to 60 nuclear interaction 
lengths. The two arrays are located 18\,m from the interaction point along the 
beam pipes downstream of the first beam-line deflecting (DX) magnet. The DX 
magnets work as sweeping magnets for the spectator protons. The FCAL measures the 
energy of the spectator protons. The resolution of the FCAL on the north side for measuring 100 GeV 
protons from the deuteron fragmentation was $\sigma=$ 40 GeV.

\subsection{Centrality analysis\label{sec:cent}} 

\begin{figure}[htb]
\includegraphics[width=1.0\linewidth]{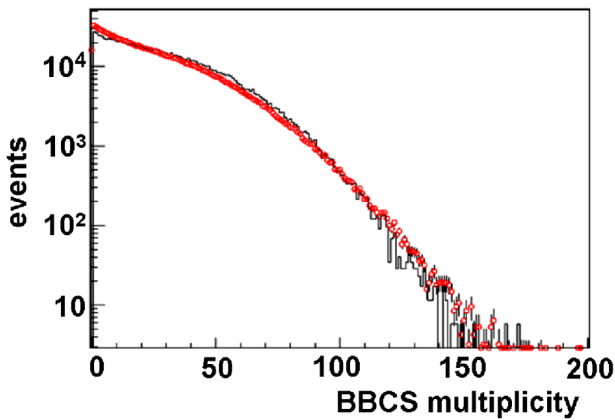}
\caption{\label{fig:bbcs_hits} (Color online)
Comparison between the experimental data hit distribution in BBCS in 
\dau\ collisions (open circles) and the calculated BBCS hit distribution 
(solid line).}
\end{figure}

The present analysis is based on minimum bias events, defined by a coincidence of 
at least one photomultiplier each in the north and south BBCs. The data were 
taken for events with vertex position within $|z|<30$\,cm along the beam axis. A 
total of 6.2 $\times 10^{7}$ events were analyzed, which corresponds to 1.6 
$\rm{nb}^{-1}$ of total integrated luminosity. At this vertex cut the minimum bias 
trigger cross section measured by BBC is 1.99 b $\pm$ 5.2$\%$~\cite{white}. Thus 
at the trigger efficiency~\cite{minbias} of 88.5$\%$ $\pm$4$\%$ we get the total 
inelastic \dau\ cross section of 2.26 $\pm$ 0.1 b. Centrality classes in 
\dau\ events were defined by charged particle multiplicity in the BBC south (BBCS) (the 
gold fragmentation side). We assume that the number of charged particles firing 
the BBCS is linearly proportional, on average, to the number of participants from 
the gold nucleus in the reaction. To check this, the BBCS response was simulated 
as a superposition of independent $N_{\rm{targ}}$ nucleon-nucleon type reactions, 
where $N_{\rm{targ}}$ is the number of participating nucleons in the struck gold 
nucleus.  As a baseline for the BBCS modeling we use unbiased data from previous 
RHIC \pp\ runs where proton-proton inelastic collisions were selected with a 
trigger synchronized to fire whenever filled bunches crossed in the PHENIX 
interaction region. $N_{\rm{targ}}$ was calculated using a Glauber 
model~\cite{minbias} with a particular parameter set~\cite{glaubparam}. The 
resulting distribution for minimum bias events is plotted in 
Fig.~\ref{fig:bbcs_hits} together with experimental data.

\begin{figure}[htb]
\includegraphics[width=1.0\linewidth]{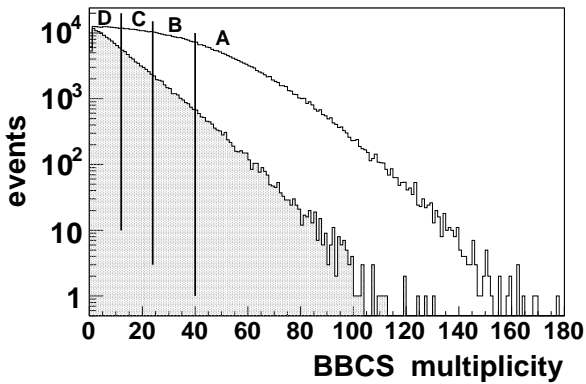}
\caption{\label{fig:bbcs} 
Multiplicity distribution in BBCS, located on the gold nucleus 
fragmentation side. Four centrality classes for \dau\ collisions are 
defined by slicing the BBCS distribution, shown with vertical lines. The 
same multiplicity cuts were used for the tagged sample of $p+$Au and 
$n+$Au events, the summed distribution of which is shown in the lower 
histogram.}
\end{figure}

\begin{figure}[htb]
\includegraphics[width=1.0\linewidth]{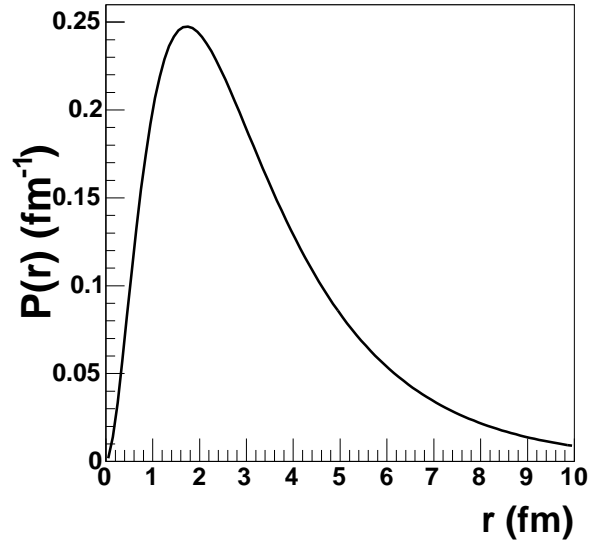}
\caption{\label{fig:d_density} 
Probability density distribution for the proton-neutron distance in the 
deuteron given by the square of the Hulth$\acute{\rm{e}}$n wave 
function~\cite{glaubparam}.}
\end{figure}

\begin{figure*}[htb]
\includegraphics[width=0.48\linewidth]{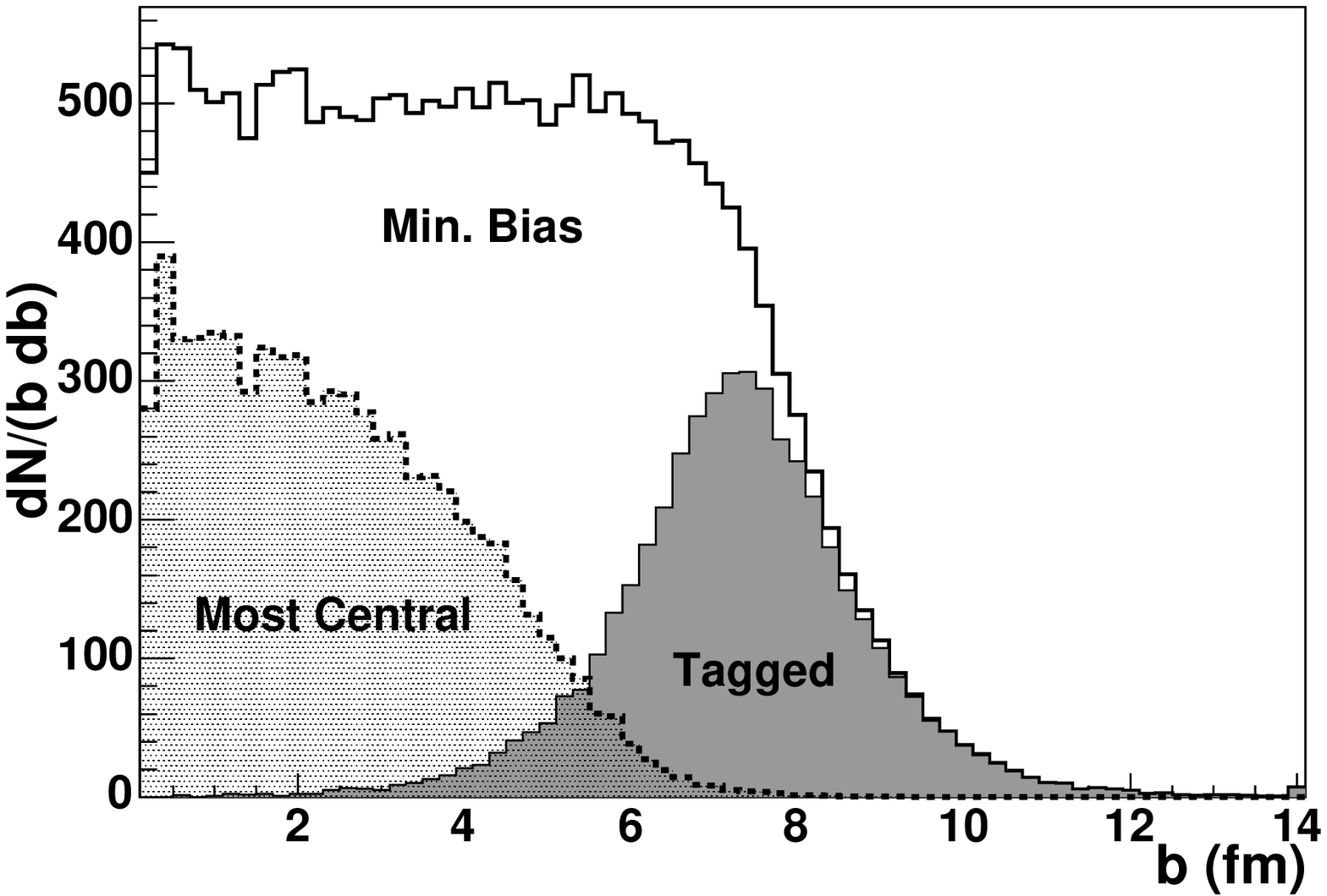}
\includegraphics[width=0.48\linewidth]{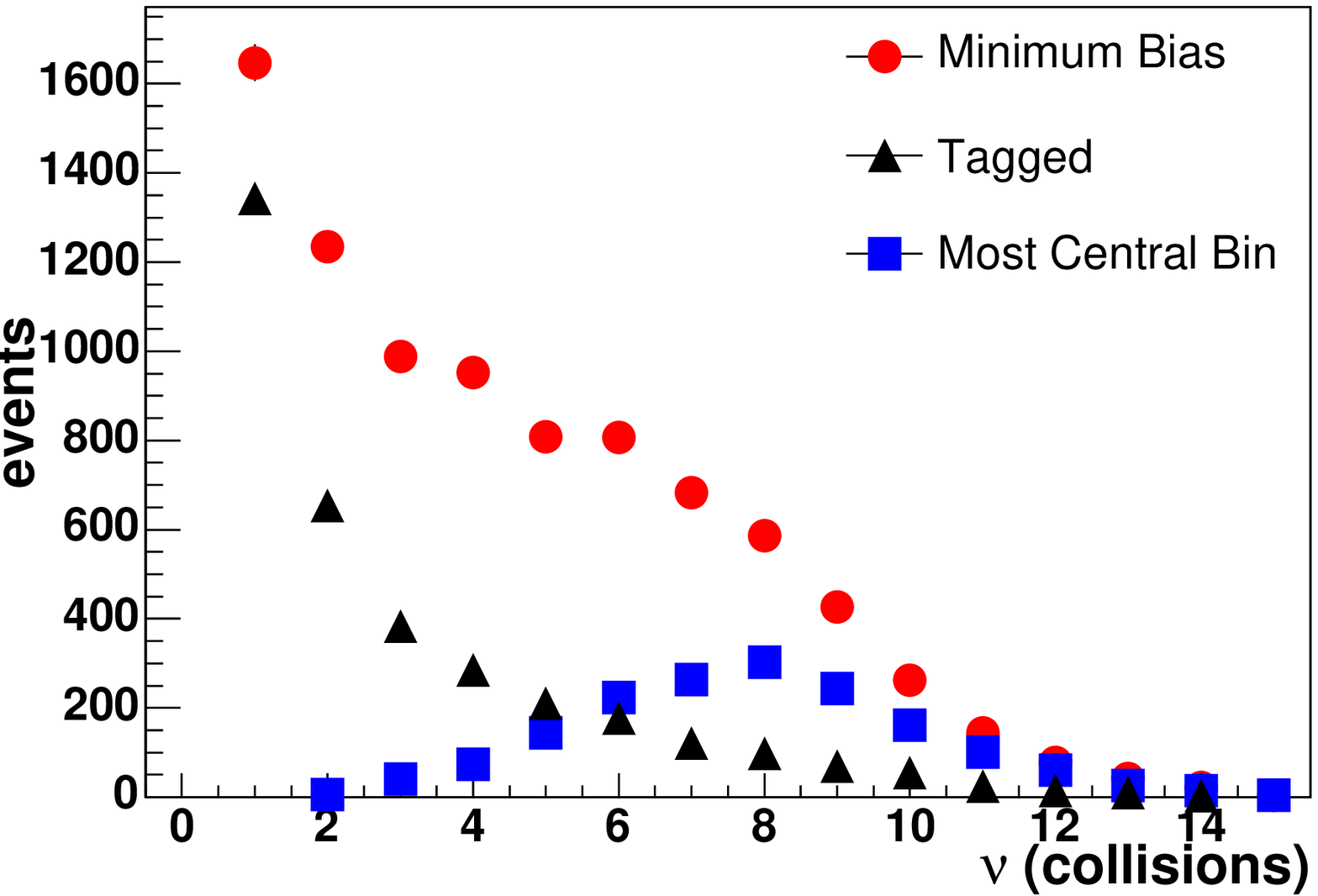}
\caption{\label{fig:b_plot} (Color online)
Left: impact parameter distribution for the minimum bias \dau\ 
collisions, for the most central events (centrality bin A) and for the 
tagged sample. For tagged events, the impact parameter was defined from 
the center of the deuteron. Right: the corresponding distribution 
of the number of collisions per participant nucleon from deuteron, 
$\nu$.}
\end{figure*}

\begin{figure}[htb]
\includegraphics[width=1.0\linewidth]{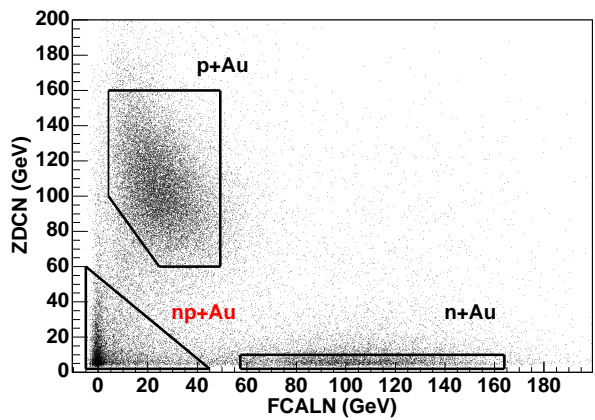}
\caption{\label{fig:fcal_zdc} (Color online)
Scatter plot of ZDCN (vertical axis) and FCALN (horizontal 
axis) signals on the deuteron fragmentation side. Solid lines show cuts which 
define the $p+$Au and $n+$Au collisions.}
\end{figure}

\begin{figure}[htb]
\includegraphics[width=0.75\linewidth]{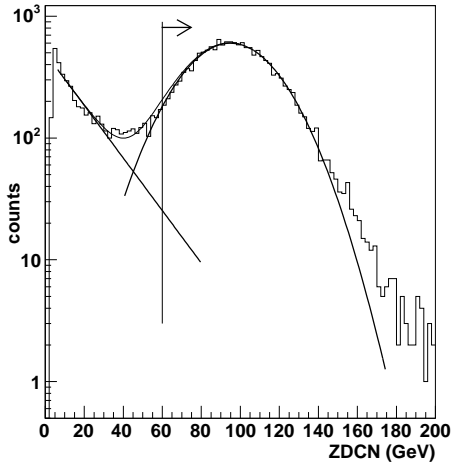}
\includegraphics[width=0.75\linewidth]{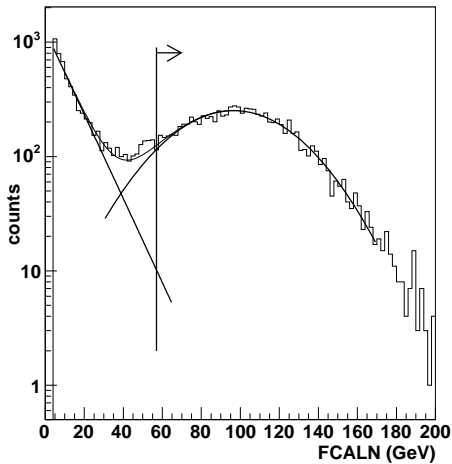}
\caption{\label{fig:tag} 
ZDCN and FCALN spectra for the most central 
tagged events. Detector response to the spectator nucleon is fit with a 
Gauss function. Background from the left side of the spectra was fit to 
an exponential function and extrapolated to the region above our 
cuts. All events to the right of the cuts, shown by vertical solid lines, 
were defined as a tag sample.}
\end{figure}

\begin{figure}[htb]
\includegraphics[width=0.75\linewidth]{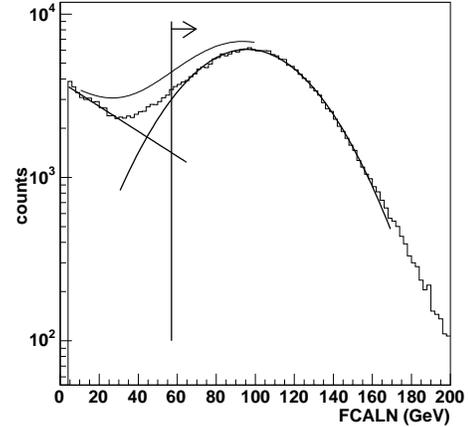}
\caption{\label{fig:fcalperiph}
FCALN spectrum for the most peripheral tagged events, with the same 
line notation as Fig.~\ref{fig:tag}. The plot illustrates the difficulty 
of evaluating the background contribution in the most peripheral 
collisions: background fit with an exponentially falling spectrum 
significantly overestimates the contribution to the events above our cut.  
}
\end{figure}

\begin{figure}[htb]
\includegraphics[width=1.0\linewidth]{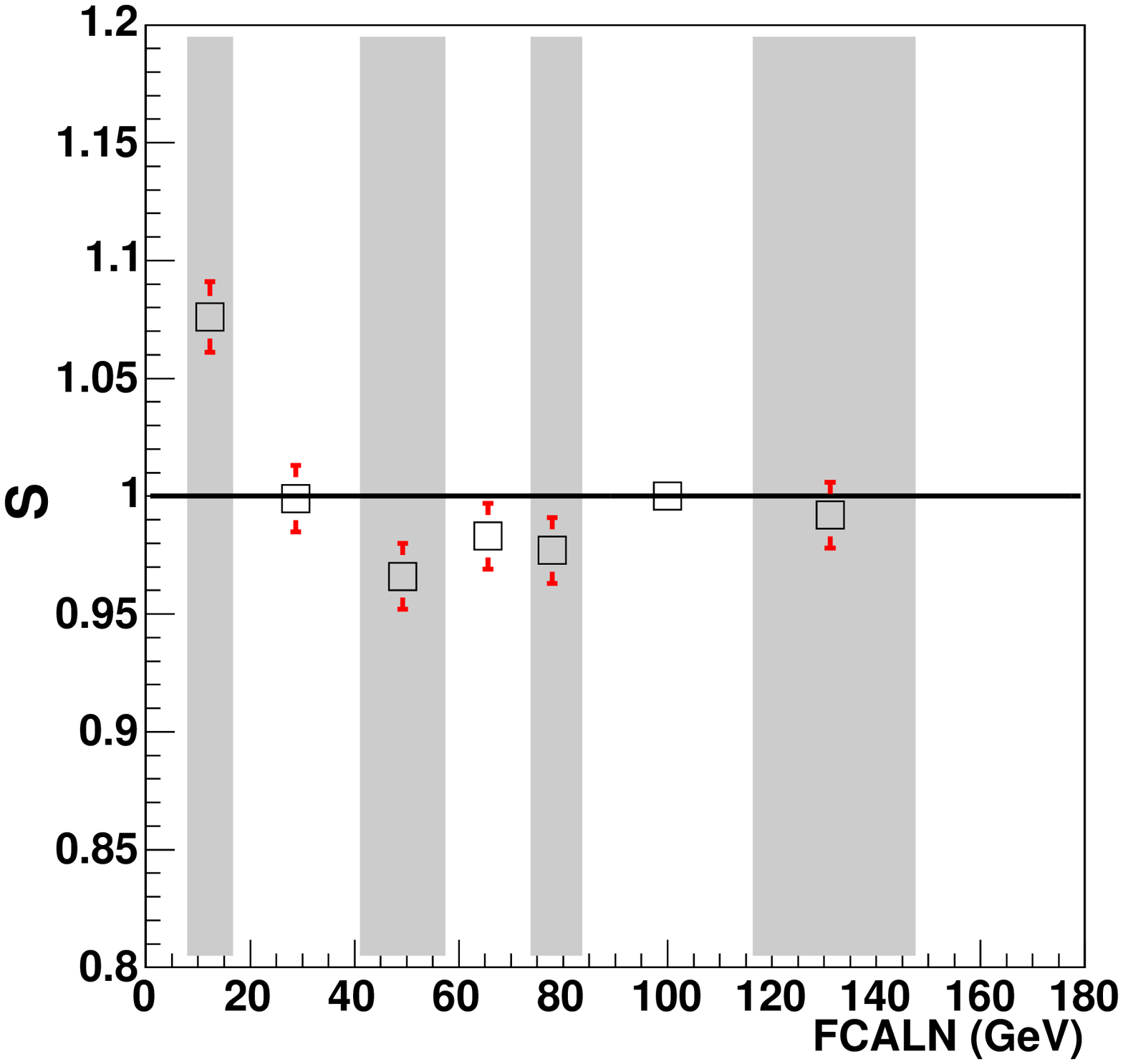}
\caption{\label{fig:stability} (Color online)
Stability of the results for different FCALN bins, see text for 
details.}
\end{figure}

The calculation appears to follow the general features of the data distribution, 
which supports our assumption that the number of BBCS firings is proportional to 
$N_{\rm{targ}}$.  We define four centrality classes for \dau\ collisions by slicing the 
BBCS multiplicity distribution into four regions, A, B, C, and D 
(Fig.~\ref{fig:bbcs}). These regions were selected to define in \dau\ collisions 
four centrality classes 0--20\%, 20--40\%, 40--60\%, and 60--88\%, respectively. 
We have taken into account the limited trigger efficiency for the peripheral 
collisions. The tagged sample, in which only one nucleon from the deuteron 
projectile interacts, has a differently shaped BBCS multiplicity distribution 
than the general data.  Nevertheless, for both the tagged sample and the general 
data the same BBCS selections were used to define the collision centrality.

In the Glauber model Monte Carlo simulation of \dau\ collisions the deuteron wave 
function was represented by the Hulth$\acute{\rm{e}}$n form \cite{glaubparam}.\footnote{For the gold nucleus, we use the Woods-Saxon density 
distribution with radius R=6.38 fm, diffuseness parameter $a$=0.54 fm and N-N cross 
section $\sigma_{NN}^{inel}$ = 42 mb. The deuteron is described by a 
Hulth$\acute{\rm{e}}$n wave function with $\alpha$ = 0.228 fm$^{-1}$ and 
$\beta$ = 1.18 fm$^{-1}$, see Ref. \cite{glaubparam}.} The 
square of this wave function determines the probability distribution for the 
proton-neutron distance in the deuteron, as shown in Fig.~\ref{fig:d_density}.  
The deuteron is a large system with a mean proton-neutron distance of about 3 fm 
and a significant probability to be larger.  In the left side of Fig.~\ref{fig:b_plot}, we 
present the calculated impact parameter distributions of \dau\ collisions and the 
tagged $N+$Au sample. On the right side of Fig.~\ref{fig:b_plot}, the corresponding 
distributions of the number of collisions per participant nucleon from the 
interacting deuteron, $\nu$, are plotted.  The parameter $\nu$ is comparable to 
the number of collisions suffered by the proton in $p+$A experiments. 
The impact 
parameter $b$ is defined as the distance between the centers of the colliding 
nuclei, Au and \textit{d}.  
This means that for the tag sample, the interaction point of participant 
nucleon from the deuteron is closer to the Au center than the distance $b$, and the spectator nucleon 
leaves the interaction region at a distance larger than $b$.  
We also did a Monte Carlo simulation with the most recent deuteron wave 
function parametrization from Ref. ~\cite{new_param}. We found that the difference 
between the calculation with the Hulth$\acute{\rm{e}}$n form and the most recent parametrization 
for the number of nucleon collisions does not exceed 1\%. 

In Fig.~\ref{fig:fcal_zdc} we illustrate our tag selection cuts. A $p+$Au 
collision event is tagged by the detection of a spectator neutron in the ZDC north 
(ZDCN) on the deuteron fragmentation side.  Similarly, we use the FCAL north 
(FCALN) on the deuteron fragmentation side to detect a spectator proton and 
thereby tag $n+$Au collision events. The scatter plot in Fig.~\ref{fig:fcal_zdc} 
shows the ZDCN and FCALN signals and has three distinct regions. Region 1 is 
defined as small or no signal in both the ZDCN and FCALN, which corresponds to 
the case in which both nucleons from the deuteron interact with the Au nucleus. Region 2 
has a small signal in the ZDCN and about 100 GeV amplitude signal in the FCALN. 
This corresponds to tagged $n+$Au collisions. Region 3 has a small signal in the 
FCALN and about 100\,GeV energy release in the ZDCN. Events in this region are 
tagged $p+$Au events. In Region 3, there is a small (anti) correlation between 
the ZDCN and FCALN. The reason for this is the close proximity of the FCALN to 
the ZDCN, see Fig.~\ref{fig:zdc_fcal}. The ZDCN effectively acts as a secondary 
target for 100\,GeV spectator neutrons. There is some contamination of secondary 
particles produced in the ZDCN into the large volume FCALN.

The purity of the tag samples from possible background contamination was 
thoroughly investigated. Figure ~\ref{fig:tag} shows ZDCN and FCALN responses for 
the most central events. Background may contribute to the tag sample from the 
left side of the spectrum with low amplitudes. To estimate the background 
contamination we fitted the left part of the spectrum using an exponentially falling 
function and fitted a Gaussian function to the detector response to the spectator 
nucleon. The thin smooth line in Fig.~\ref{fig:tag} shows the sum of the two 
functions. The background under the Gaussian peak was estimated as the integral 
of the exponential function above our cut, which is marked by the vertical line 
and arrow. In central $p+$Au events we estimate 2.8\% contamination in the ZDCN 
spectrum. For more peripheral events this contamination decreases and reaches 
1.4\% in the most peripheral bin.

The background contamination in the FCALN defined $n+$Au sample is more 
complicated. For the most central events, as shown in Fig.~\ref{fig:tag} on the 
right, the background contributes 0.4\% to the area above our cut. As centrality 
changes so too does the shape of the background spectrum.  
Figure ~\ref{fig:fcalperiph} illustrates the problem of background estimation. 
Attempts to fit the background spectrum with an exponential function, as was done 
for central events, failed. The sum of the estimated background and the Gauss fit 
is well above the experimental data implying that the background spectrum falls 
much faster than a simple exponential function. However, the relevant question in 
assessing the background contamination is not its absolute or relative value, but 
rather how much it could distort our experimental data. 
The goal of this analysis is to measure the charged hadron spectrum in the PHENIX 
central arms. Thus, we can estimate the stability of our results measured by the 
central arms versus different cuts applied to the FCALN. We define the stability 
of the measurement, $S$, in a particular FCALN bin (Fig.~\ref{fig:stability}) as 
the ratio of the number of charged particle tracks per event with \pt\ $>$ 
0.5\,GeV/$c$ for this cut to a reference cut in which the signal to background 
ratio is the largest.  For the reference cut we use 100 $\pm$ 20 GeV FCALN response 
(within region 3 as defined above).  This is the most probable detector response 
for a spectator proton as seen in Fig.~\ref{fig:fcalperiph}.  We observe that in 
the region just to the left of our reference cut, where we may expect about equal 
amount of signal and background events, the parameter $S$ is only 3\% smaller 
than unity, as seen in Fig.~\ref{fig:stability}. Within errors, $S$ does not 
depend on track momentum. Thus, in the worst case the background contamination 
effect is less than 2\% for events above our cut in the $n+$Au tagged sample.


We do not correct the tagged sample results for any contamination.
We average the two tagged samples to form a single nucleon + Au 
($N$$+$Au) sample.

For each of the four BBCS multiplicity regions shown in 
Fig.~\ref{fig:bbcs}, using a Monte Carlo model within the Glauber 
multiple collision formalism~\cite{glauber}, we calculate for the 
\dau\ and the tagged $N$$+$Au collisions the average number of participating 
nucleons $N_{\rm{part}}$, the nuclear overlap function $T_{AB}$, the number of 
collisions $N_{\rm{coll}}$, and number of collisions per participant nucleon 
from the deuteron $\nu$. The calculated values are presented in 
Table~\ref{tablenn}.  The nuclear overlap function $T_{AB}$ is defined 
as
\begin{equation}
T_{AB}(b)=\int d^2\,\vec s \,T_A(\vec s)\,T_B(|\vec b-\vec s|),
\label{eq:TAB_defined}
\end{equation}  
where the integration is performed over the element of overlapping area 
$d^2\,\vec s$, and $\vec s=(x,\,y)$ is a vector in the transverse plane of 
interacting nuclei at the impact parameter $\vec b$ between the centers 
of the nuclei. For nucleus A the nuclear thickness function $T_A(b)$ 
is defined as
\begin{equation}
T_{A}(b)=\int dz\,\rho_A(b,z). 
\label{eq:TA_defined}
\end{equation}

\begingroup \squeezetable
\begin{table}[htb]
\caption{\label{tablenn} 
Total number of participants $N_{\rm{part}}$, number of collisions $N_{\rm{coll}}$, 
nuclear overlap function $T_{AB}$, see Eq. ~\ref{eq:RAA_defined}, 
average number of collisions per participant nucleon from deuteron $\nu$, 
and the BBC bin correction factor for different centrality classes.}
\begin{ruledtabular}
\begin{tabular}{cccccc}
  Cent. bin &$ \langle N_{\rm{part}}\rangle$   & $\langle N_{\rm{coll}}\rangle$      & $ \langle T_{AB}\rangle$, $\rm{mb}^{-1}$    & $\nu$ &$C_{\rm{BBC}}$ \\ \hline
A   &15.0$\pm$1.0   &15.4$\pm$1.0    & 0.37$\pm$0.02   &7.5$\pm$0.5 & 0.95$\pm$0.03\\
B  &10.4$\pm$0.4   &10.6$\pm$0.7  & 0.25$\pm$0.02 &5.6$\pm$0.4 & 0.99$\pm$0.01\\
C  &7.0$\pm$0.6    &7.0$\pm$0.6   & 0.17$\pm$0.01   &4.0$\pm$0.3 & 1.03$\pm$0.01\\
D  &3.2$\pm$0.3    &3.1$\pm$0.3   & 0.07$\pm$0.01  &2.2$\pm$0.2 & 1.04$\pm$0.03\\  
\hline
tag A  &10.6$\pm$0.7    &9.6$\pm$0.7  & 0.23$\pm$0.02    &9.6$\pm$0.7  & 0.93$\pm$0.03 \\
tag B &8.0$\pm$0.6     &7.0$\pm$0.6  & 0.17$\pm$0.02    &7.0$\pm$0.6  & 0.95$\pm$0.02 \\
tag C &5.6$\pm$0.3     &4.6 $\pm$0.3 & 0.11$\pm$0.01    &4.6$\pm$0.3 & 0.95$\pm$0.02 \\
tag D &3.1$\pm$0.2	   &2.1$\pm$0.2  & 0.05$\pm$0.01     &2.1$\pm$0.2  & 0.97$\pm$0.04 \\  
\end{tabular}
\end{ruledtabular}
\end{table}
\endgroup 

Normalization of $T_{AB}(b)$ is done by integration over all impact 
parameters:
\begin{equation}
\int d^2\,b\,T_{AB}(b)=A\,B .
\label{eq:TAB_norm1}
\end{equation}
The average number of binary inelastic nucleon-nucleon collisions at 
impact parameter $b$ was calculated from $T_{AB}(b)$ as
\begin{equation}
\langle N_{coll}\rangle = \sigma _{NN} \, T_{AB}(b), 
\label{eq:TAB_norm2}
\end{equation}
where $\sigma_{NN}$ is the inelastic nucleon-nucleon cross 
section~\cite{glaubparam}.

An additional multiplicative correction factor, $C_{\rm{BBC}}$, see 
Table~\ref{tablenn}, has been applied to the data for different 
centrality bin selections~\cite{nagle}. This correction addresses two 
effects, each of which distorts the centrality classification in the 
opposite direction. Because of natural fluctuations in the number of 
produced charged particles at a particular impact parameter, the BBCS 
centrality selections have imperfect resolution. In the case of a steeply 
falling BBCS multiplicity spectrum, especially for the tagged sample (see 
Fig.~\ref{fig:bbcs}), there is a contamination of peripheral collision 
events into a more central event class. This effectively decreases the 
actual number of $N_{\rm{coll}}$ and particle production in the central 
events. The second effect is due to the BBC coincidence required for a 
PHENIX event trigger. In all calculations we used for the inelastic 
nucleon-nucleon cross section $\sigma _{NN}$ = 42 mb. Actually, this cross 
section has three components: nondiffractive collisions with 28 mb, 
single diffractive collisions with 10 mb, and double diffractive 
collisions with 4 mb cross section~\cite{pythia}. From Monte Carlo 
simulation of \textit{p}+\textit{p} collisions we found BBC trigger efficiencies of about 
(72 $\pm$ 1)\%, (7 $\pm$ 1)\% and (32 $\pm$ 1)\%, for each process 
respectively. In the Monte Carlo simulation we use the PYTHIA 6.2 event 
generator~\cite{pythia} and the GEANT simulation of the BBC detector.  
Single and double diffractive collisions produce particles dominantly 
near the beam rapidity and thus have a small probability for particle 
production in the BBC acceptance of 3.0$< |\eta| <$ 3.9, and an even 
smaller probability at midrapidity. Therefore, the BBC trigger is biased 
to the nondiffractive collisions, which have larger particle production 
at midrapidity.  In more central events as the number of 
nucleon-nucleon collisions increases the probability that there will be 
at least one nondiffractive collision approaches 100\%; thus the BBC bias 
becomes negligible in central events.

\subsection{Charged hadron analysis\label{sec:charged}}

The present analysis follows the methods of the analysis described in 
Ref. ~\cite{ppg023}.  The majority of background tracks are particles 
with low momenta, which in traveling from DC to PC3 undergo multiple 
scattering and are additionally deflected in the residual magnetic field 
behind the DC.

\begin{figure*}[hbt]
\includegraphics[width=0.48\linewidth]{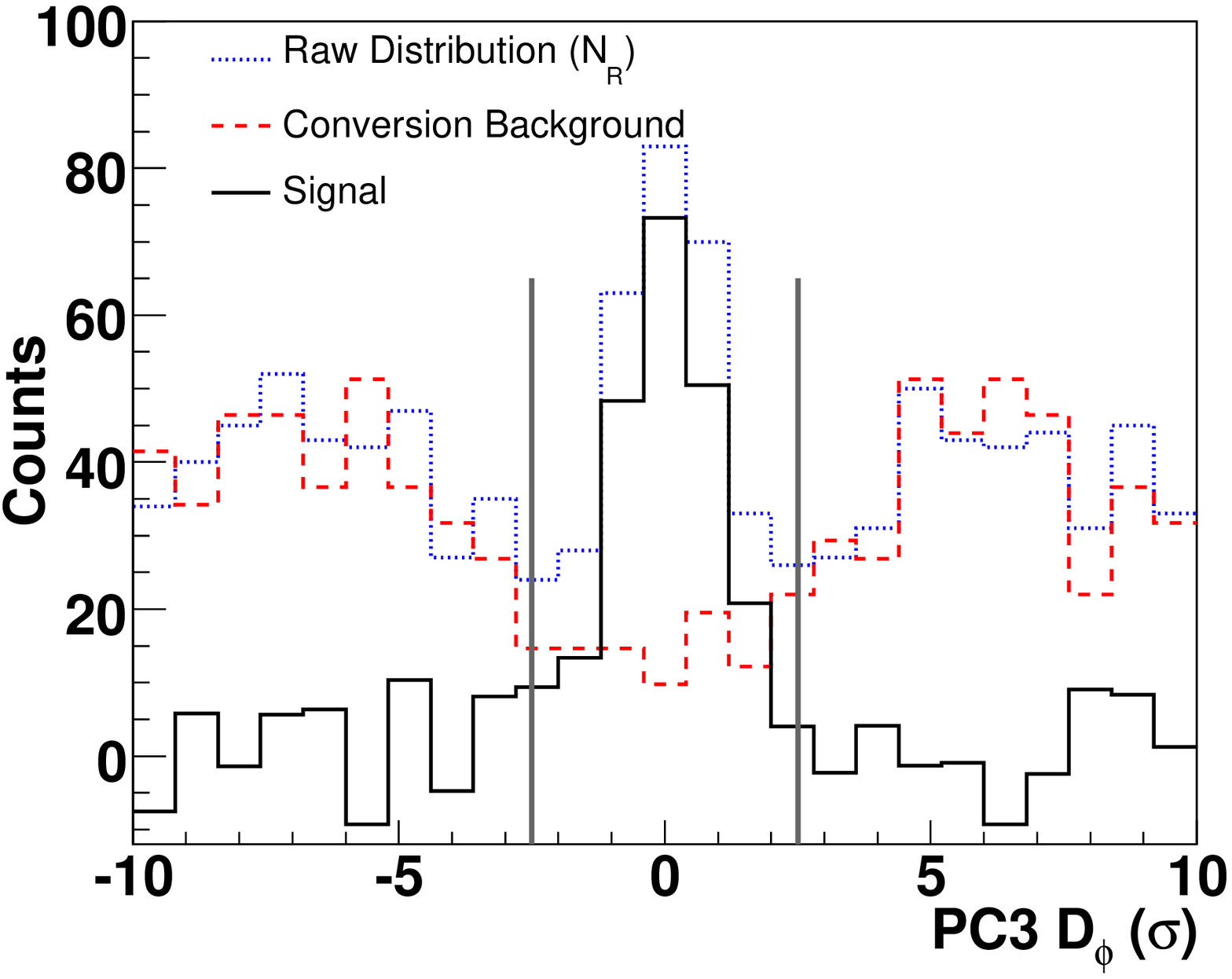} 
\includegraphics[width=0.48\linewidth]{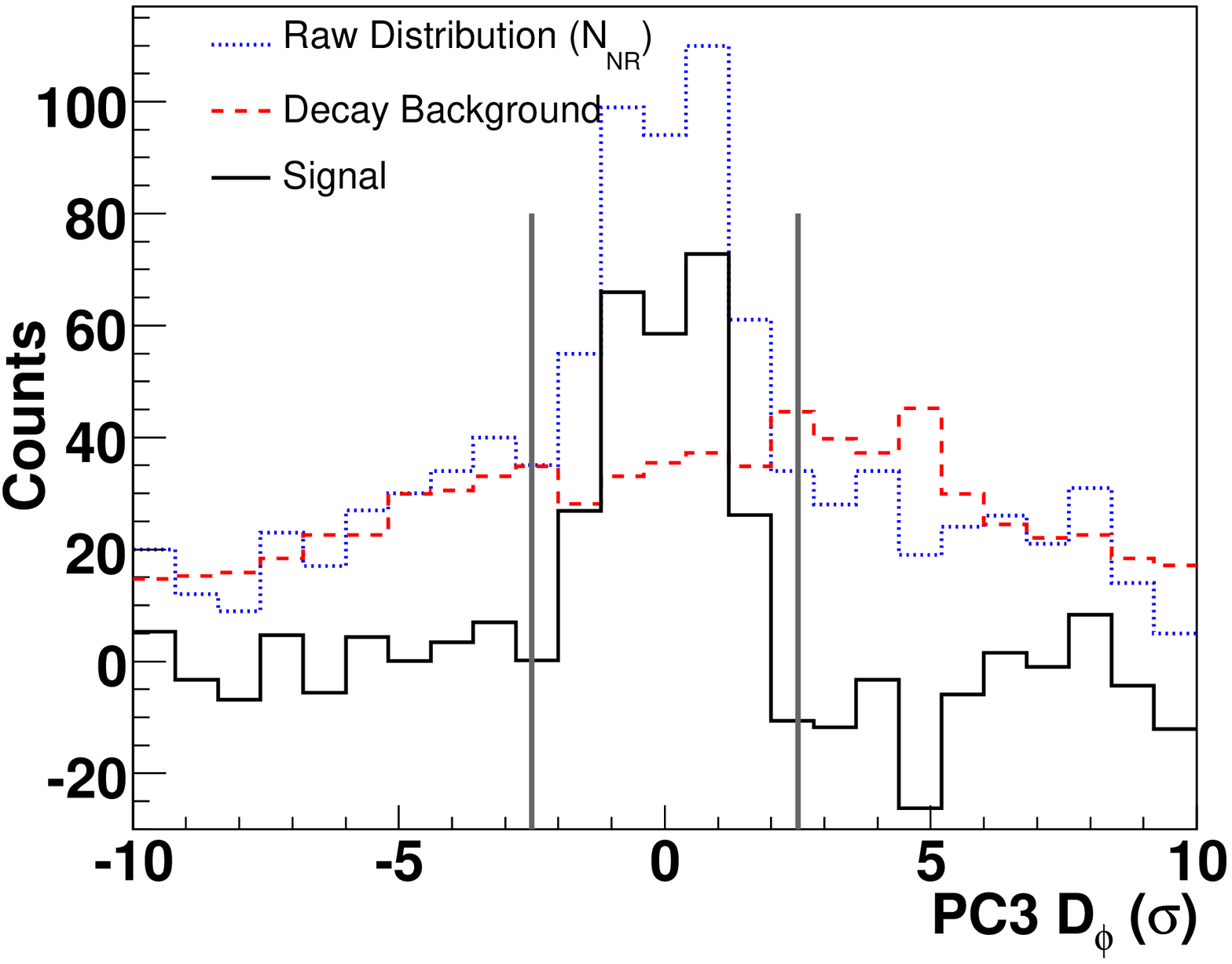} 
\caption{\label{fig:PC3} (Color online)
PC3 $D_{\phi}$ distributions for the 
conversion subtraction (left) and the decay subtraction (right) shown for 
minimum bias with $6<p_{T}<7$ \gevc. The raw distributions, $N_{R}$ and 
$N_{NR}$ for conversion and decay respectively, are shown with dotted 
lines.  The estimated conversion background shown with a dashed line is 
$N_{e}/R_{e}$.  The estimated decay background, shown with a dashed line, 
was obtained by scaling the PC3 distribution of $N_{NR}$ tracks with 
$p_{T}>10.5$ \gevc based on the ($4\sigma<\vert D_{\phi} 
\vert < 10\sigma$) region.  In both plots, the signal is the raw minus the 
estimated background distributions.  The vertical bars show the track 
matching cuts at $\pm$ 2.5 $D_{\phi}$. }
\end{figure*}

To minimize this background we employ a track matching cut in the PC3 
that rejects tracks whose displacement in the $\phi$ or $z$ direction, 
$D_{\phi}$ and $D_{z}$, respectively, is greater than 2.5 standard 
deviations.  In addition we make a fiducial cut around the $z$ vertex 
determined by the BBC. Despite these veto cuts, there is still 
significant background contamination for \pt\ above 4 \gevc which must be 
subtracted.  The main sources of the remaining background are $e^+e^-$ 
pairs from photon conversions in the material between the collision 
vertex and the drift chamber, and secondary particles produced by hadron 
decays.

To distinguish these backgrounds, we use the RICH detector to divide all 
tracks into two subsets: tracks with an associated RICH signal, $N_{R}$, 
and tracks with no signal in the RICH, $N_{NR}$. Tracks with at least one 
hit in the RICH contain high \pt\ pions and conversion electrons. For 
reconstructed electrons with momentum above 150\,MeV/$c$, the average 
number of photomultiplier tube (PMT) hits in the RICH associated with the 
track is $\langle N_{\rm{PMT}} \rangle \approx$ 4.5.

The RICH detects more than 99\% of all conversion electrons for 
$N_{\rm{PMT}}\ge 1$. At this threshold the RICH also detects pions with 
\pt$\ge 4.8$\,GeV/$c$, but the number of associated PMTs for pions 
reaches its asymptotic value well above 10\,GeV/$c$; for a 10\,\gevc\ pion 
$\langle N_{\rm{PMT}} \rangle = 3.6$.  Therefore we label tracks with 
$N_{\rm{PMT}}\ge$ 5 as electron tracks, ${N_e}$, which compose some fraction, 
$R_{e}$, of conversion electrons.  To calculate this fraction, we take 
advantage of the deflection of conversion electrons in the magnetic field 
between the DC and PC3. This deflection leads to poor track matching in 
PC3 which distinguishes electrons from true high \pt\ pions.  We define 
poor PC3 track matching as a displacement of more than four standard 
deviations in the $\phi$ direction.  Thus we measure the value of $R_{e}$ 
as the fraction of $N_{R}$ with 4$\sigma<\vert D_{\phi}\vert<10\sigma$.  
For minimum bias events we find $R_{e} = 0.41 \pm 0.01$.  The real pion 
signal $S_R$ in the $N_{R}$ sample, is calculated for each \pt\ bin as
\begin{equation}  
S_R =N_{R}-\frac{N_e} {R_e}
\end{equation}
The PC3 distribution for the conversion subtraction is shown on the left 
side of Fig.~\ref{fig:PC3}. The conversion subtraction is performed 
independently in each centrality bin.  The definition of ${N_e}$ (tracks 
with $N_{\rm{PMT}}\ge 5$) does not perfectly select electron tracks, as some 
fraction of pions satisfies the cut.  This leads to a fraction of 
authentic pions, which have $N_{\rm{PMT}} \ge 5$, being subtracted along with 
the conversion electrons. This fraction is small below 7\gevc, but 
increases rapidly for higher $p_{T}$. We calculate a correction 
factor to address this over-subtraction, using a Monte Carlo simulation 
of the detector.
The uncertainty associated with this correction, 
$\delta_{\pi \ \rm{loss}}$, is shown later in Table~\ref{systerrpt}.

\begin{figure}[htb]
\includegraphics[width=1.0\linewidth]{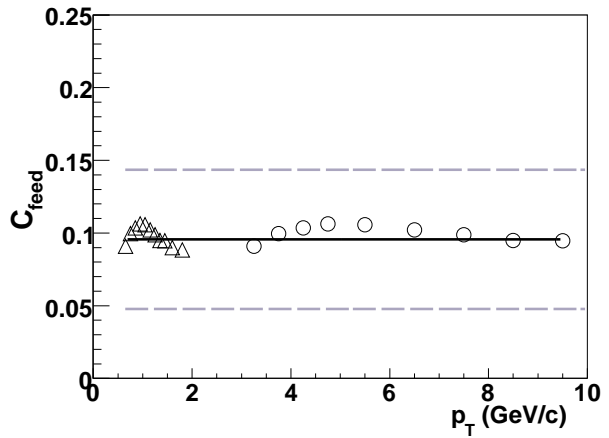}
\caption{\label{fig:feed} 
Feed-down contamination as the ratio of protons and charged pions from 
feed-down decays to total detected charged hadrons.  The low \pt\ points 
shown with triangles are calculated using the fraction of $p + K$ 
particles to hadrons measured in Ref. ~\cite{felix}.  The higher \pt\ points are 
calculated using the various ratios of Eq. ~\ref{eq:feedeq}.  The 
solid line is a fit to both sets of points and is used as $C_{\rm{feed}}$.  
It is bracketed by dashed lines showing the assigned 50\% systematic 
uncertainty. }
\end{figure}

\begin{figure*}[hbt] 
\includegraphics[width=1.0\linewidth]{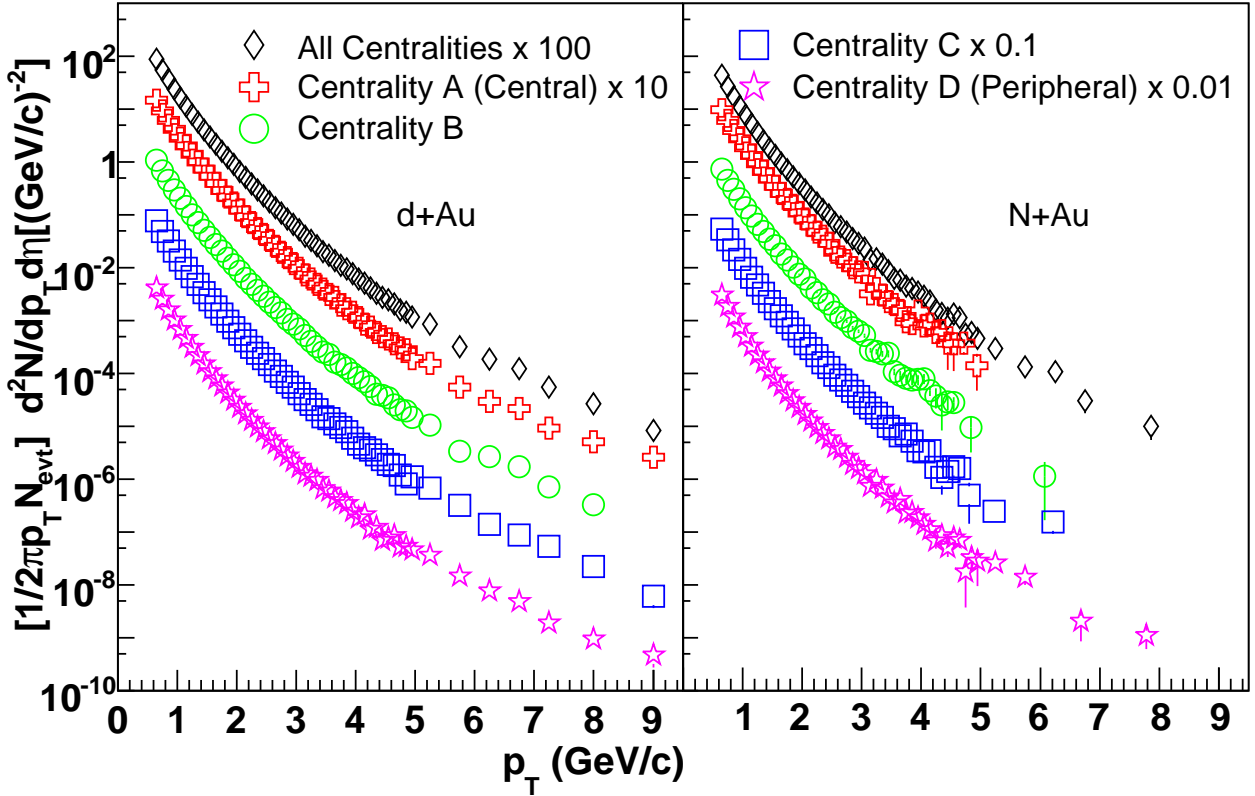} 
\caption{\label{fig:data} (Color online)
Charged hadron spectra for centrality selected $d+$Au collisions and $N+$Au collisions.  The error bars 
represent statistical uncertainties only, although they are generally 
smaller than the data point symbols.  Points from different centrality 
bins are scaled sequentially by a factor of 10.  (See 
Table~\ref{tablenn} for centrality class definitions.)}
\end{figure*}

\begin{figure}[htb]
\includegraphics[width=1.0\linewidth]{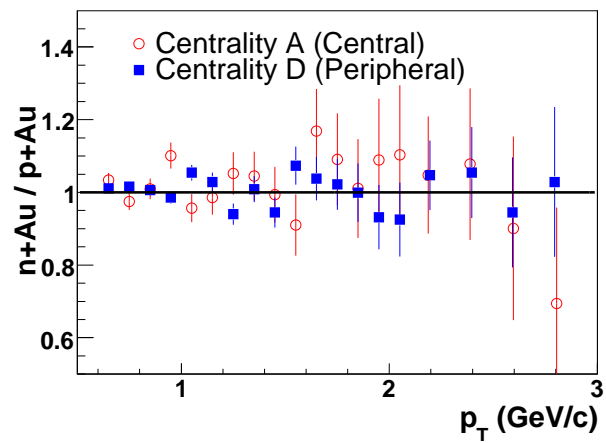}
\caption{\label{fig:ptag_ntag} (Color online)
Ratio of $n+$Au to $p+$Au invariant yield per event shown in the most 
central and peripheral bins.  The error bars are statistical only, as all 
systematic uncertainties cancel in the ratio.(See 
Table~\protect\ref{tablenn} for centrality class definitions.)}
\end{figure}

The $N+$Au sample is averaged over $p+$Au and $n+$Au events. This is 
based on the good agreement between the yields of the two tagged samples, 
as shown in Fig.~\ref{fig:ptag_ntag}.

The hadron decay background is of two types: ``decay" and ``feed down.''  
The decay background is produced by $\pi$ and $K$ decays far from the 
source and thus with reconstructed momenta different from their true 
momenta, whereas the feed-down background is produced by weak decays of 
short-lived particles, mostly $K^0_{s}$ and $\Lambda$ particles near the 
event vertex with apparent momenta close to their true momenta.

From tracks with no RICH signal, the $N_{NR}$ sample, we define a 
subsample by selecting tracks with \pt\ $> 10.5$\,GeV/$c$, a \pt\ region 
which is almost exclusively background.  We expect that its shape in 
$D_{\phi}$ will be the same as the background in the lower \pt\ region.  
Within this subsample we calculate the ratio, $R_{\rm{decay}}$, of tracks 
which pass the PC3 cut ($\vert D_{\phi}\vert<2.5\sigma$) to those with a 
poor match ($4\sigma<\vert D_{\phi} \vert <10\sigma$). For minimum bias 
events $R_{\rm{decay}} = 0.55 \pm 0.03$. For each momentum bin, the total 
decay background is then obtained by multiplying the $N_{NR}$ tracks with 
poor PC3 matching ($4\sigma<\vert D_{\phi} \vert <10\sigma$) by 
$R_{\rm{decay}}$. The PC3 distribution for the decay subtraction is shown on 
the right side of Fig.~\ref{fig:PC3}.  The decay background as a function 
of \pt\ is measured and subtracted independently in each centrality bin.

The feed-down subtraction addresses the detected $\pi$ and $p$ particles 
that were produced in the decays of $K^0_{s}$ and $\Lambda$ particles, so 
we define the total feed-down contamination as:
\begin{equation}
C_{\rm{feed}} = \frac{(p + \pi)^{\rm{feed}}} {h^{\rm{detected}}}
\end{equation}
averaging over charged particle and antiparticle yields. For feed-down 
estimation, we have no statistical recourse and must resort to simulation 
to find the contamination.  We assume that the spectral shapes of the 
$K^0_{s}$ and $\Lambda$ follow the shapes of the charged kaon and proton 
spectra, respectively.  There is good agreement of Monte Carlo 
simulations in $d+$Au~\cite{felix} and minimum bias Au+Au~\cite{ppg023} 
for proton feed down.  We therefore use the Au+Au Monte Carlo simulation, 
which also includes $K^0_{s}$ to pion processes, to obtain the ratio 
$(p + \pi)^{\rm{feed}} / (p + K)^{\rm{detected}} = 0.2$, after the decay background 
subtraction. To make use of this ratio we rewrite the contamination as:
\begin{equation}
C_{\rm{feed}} = \frac{(p + \pi)^{\rm{feed}}} {(p + K)^{\rm{detected}}}
{\frac{(p + K)^{\rm{detected}}}{h^{\rm{detected}}}}
\end{equation}
To find the contamination we use the fraction of $p + K$ particles in our 
measured hadron sample. To do so, for $p_T$ less than 2.5\,\gevc\, we use 
the PHENIX published data on identified hadron production in 
$d+$Au~\cite{felix}.  We explicitly calculate the $(p + 
K)^{\rm{detected}}$/$h^{\rm{detected}}$ ratio from those data.  At higher \pt\, 
assuming \pizero\ has the same yield as $\pi^{\pm}$, we subdivide this 
ratio as
\begin{equation}
\label{eq:feedeq}
\frac{p + K}{h}=\frac{p + K}{\pi} 
\frac{\pi}{h}=\left( \frac{p}{\pi} + \frac{K}{\pi^{0}} \right) \frac{\pi^{0}}{h}
\end{equation}
so that the right-hand side consists of all measured quantities. The 
$p/\pi$ ratio is taken from STAR measurements~\cite{star} scaled to match 
the PHENIX data~\cite{felix} in their common \pt\ region.  The $K/\pi^{0}$ 
ratio is calculated from Refs. ~\cite{minbias} and~\cite{an421}.  The ratio of 
$\pi^{0}$s to hadrons is calculated from the charged hadron measurement 
of this analysis and \pizero\ measurements in Ref. ~\cite{minbias}.  From both 
the low and high \pt\ regions $C_{\rm{feed}}$ is calculated to be 9.6\%. To 
this factor we assign a 50\% systematic uncertainty based on uncertainty 
of the various particle ratios and the Monte Carlo simulation.  
$C_{\rm{feed}}$ is shown in Fig.~\ref{fig:feed}.  We correct for the effects 
of feed-down decay by multiplying the spectra remaining after the 
conversion and decay subtraction by $1 - C_{\rm{feed}}$.

Following the background subtraction, we constructed a single, 
\pt\ dependent correction function to correct the hadron spectra for 
acceptance, decay in flight, reconstruction efficiency, and momentum 
resolution.  The correction is determined by using a Monte Carlo 
simulation~\cite{MC} of the PHENIX detector.  The correction function is 
necessarily particle species dependent to take into account multiple 
scattering and decays, therefore we calculate separate correction 
functions for $\pi^{+}$, $\pi^{-}$, $K^{+}$, $K^{-}$, $p^{+}$, and 
$p^{-}$.  The individual functions are weighted by the particle 
\pt\ spectra measured in peripheral Au+Au collisions ~\cite{ppg26} to form a 
single correction factor, $C_{\rm{MC}}(p_{T})$, subject to a systematic 
uncertainty, $\delta_{\rm{MC weight}}$, stemming from uncertainty on the 
particle mixture~\cite{minbias}.  For absolute normalization of the 
spectra we match the geometrical acceptance of the Monte Carlo simulation 
with the actual acceptance of the data.  To obtain the charged hadron 
yield we multiply the background-subtracted spectra by the correction 
function. We normalize each centrality bin by dividing by the number of 
events, and each momentum bin by dividing by its bin width.  Each data 
point is corrected so its value corresponds to the bin center.  We define 
the invariant hadron yield as
\begin{equation}
\begin{split}
\frac{1}{N_{\rm{evt}}}\frac{d^{2}N}{2\pi p_{T} \ dp_{T} \ d\eta}= 
\left( \frac{d^{2}N}{2\pi p_{T} \ dp_{T} \ d\eta}\right) ^{\rm{bkg-subtracted}} \\
\ \frac{1}{N_{evt}} (1-C_{\rm{feed}}) \ C_{MC}(p_{T}) \ C_{BBC}
\end{split}
\end{equation}
All of the preceding steps are applied to the tagged $N+$Au (nucleon + 
gold) sample as well as to the general $d+$Au sample.

\begin{figure*}[htb]
\includegraphics[width=0.9\linewidth]{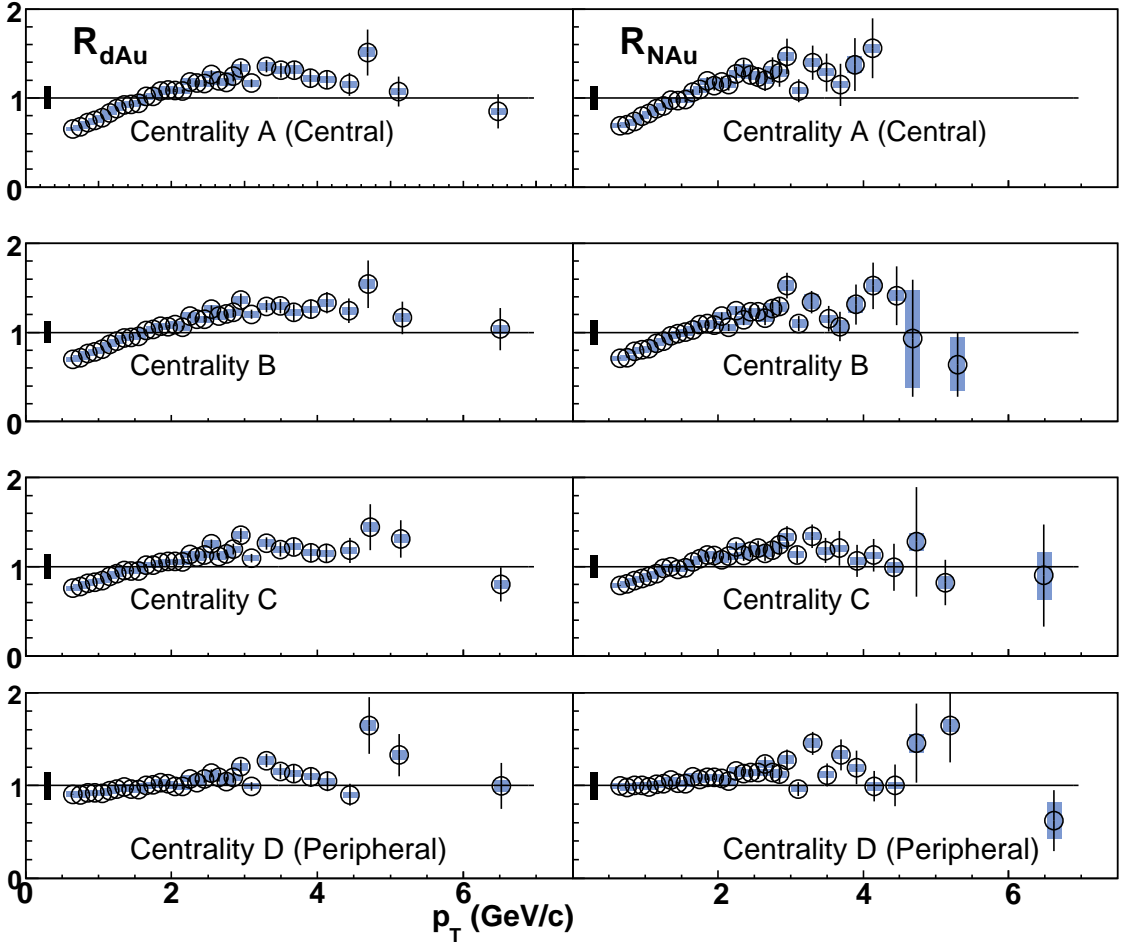}
\caption{\label{fig:RAB} (Color online)
From left to right: $R_{d\rm{Au}}$, $R_{N\rm{Au}}$ as functions of $p_{T}$.  
The bars represent statistical uncertainty (often smaller than the data 
point symbols), the shaded boxes on each point represent systematic 
uncertainties that change with $p_{T}$, and the black boxes on the 
left show systematic uncertainties that do not change with $p_{T}$. (See 
Table~\ref{tablenn} for centrality class definitions.)
} 
\end{figure*}

\begin{figure}[htb]
\includegraphics[width=1.0\linewidth]{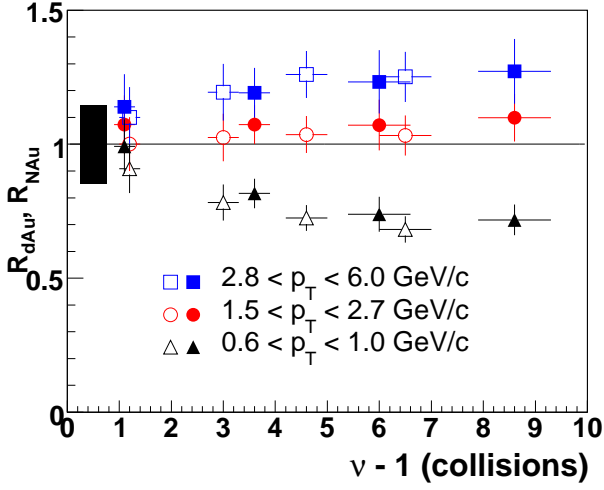}
\caption{\label{fig:nu}  Ccolor online)
$R_{d\rm{Au}}$ (open symbols) and $R_{N\rm{Au}}$ (closed symbols) values averaged in three momentum ranges as 
functions of $\nu-1$. Point-by-point uncertainty bars show the quadratic sum of the statistical 
uncertainty in $R_{AB}$ and the systematic uncertainties that change with 
$\nu$.  The box on the far left represents the size of the systematic 
uncertainty that does not change with $\nu$.  Horizontal bars show the 
uncertainty in the value of $\nu$ for each centrality class.}
\end{figure}

\begin{figure}[htb]
\includegraphics[width=0.9\linewidth]{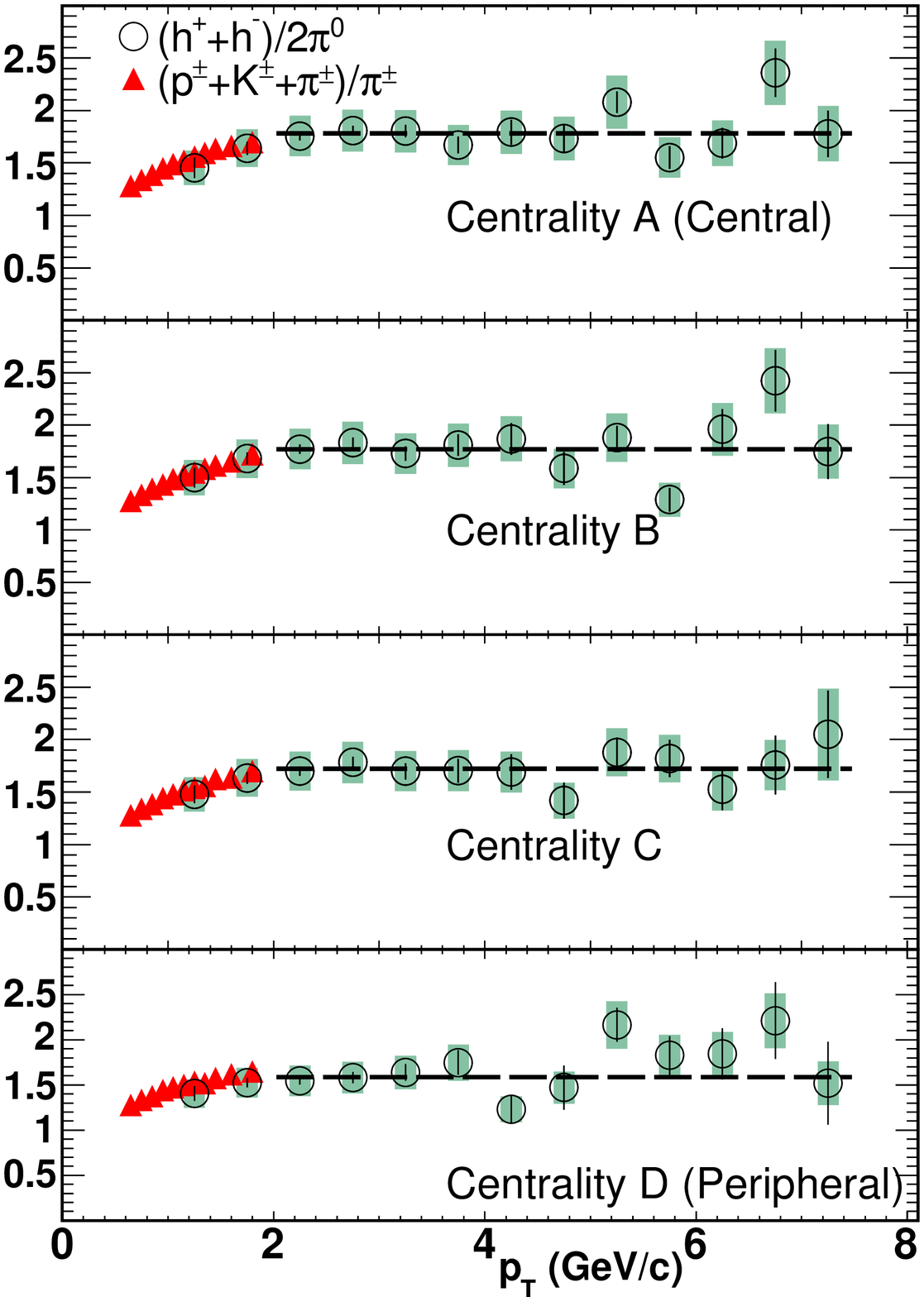}
\caption{\label{fig:pihad}  (Color online)
$(h^{+}+h^{-})/2\pi^{0}$ ratios as functions of transverse momentum from 
$d+$Au collisions in four centrality bins. Open circles are the charged 
hadron spectra from the present analysis divided by \pizero\ data 
from Ref. ~\cite{ppg044}.  Bars indicate statistical uncertainties, and the 
shaded boxes systematic uncertainties. Triangles are the 
$(p^{\pm}+K^{\pm}+\pi^{\pm})/\pi^{\pm}$ ratios from Ref. ~\cite{felix}, with 
statistical uncertainties roughly the size of the symbols. (See 
Table~\ref{tablenn} for centrality class definitions.)}
\end{figure}

%
We further examine the $d$+Au invariant yields by comparing them with the 
invariant yield from $p+p$ collisions.  Previous experiments have 
demonstrated suppression of charged hadron and pion yields in Au+Au 
collisions~\cite{ppg023}; such suppression can be quantified by the 
nuclear modification factor $R_{AB}$.  For any collision of nuclei $A+B$, 
$R_{AB}$ is calculated for each centrality class as the ratio of the 
yield in $A+B$ collision to the cross section in $p+p$ collisions scaled 
by the nuclear overlap function $\langle T_{AB}\rangle$, that is,
\begin{equation}
R_{AB}(p_T) = \frac{(1/N_{\rm{evt}}) \; d^{2}N^{A+B}/dp_T \ d\eta }
{\langle T_{AB}\rangle \; d^{2}\sigma^{p+p}/dp_T \ d\eta}. 
\label{eq:RAA_defined}
\end{equation} 
$\langle T_{AB}\rangle$ is determined by the density distribution in the 
nuclei $A$ and $B$ and is averaged over the impact parameter range within 
a particular centrality class.  In our case, nucleus $A$ refers to the 
deuteron (or the single nucleon in tagged events), and nucleus $B$ refers 
to the gold nucleus.  $\langle T_{AB}\rangle$ values are presented in 
Table~\ref{tablenn}.  Using the PHENIX $p+p$ spectra~\cite{ppref}, we thus 
calculate $R_{d\rm{Au}}$ and $R_{N\rm{Au}}$.

\begin{table}[htb]
\caption{\label{systerr} 
Systematic uncertainties that are constant for all $p_{T}$.}
\begin{ruledtabular}
\begin{tabular}{lc} 
Source & Uncertainty(\%) \\ \hline
Geometric acceptance correction & 2.9\\
Track matching & 2.2\\
Run-by-run variation & 5\\
Feed-down correction & 4.8\\
$\delta_{\rm{MC weight}}$ & 3.7\\  
Total  & 8.7\\
\end{tabular}
\end{ruledtabular}
\end{table}
\begin{table}[hbt]
\caption{\label{systerrpt} 
Systematic uncertainties that vary with $p_{T}$.  Background 
subtraction (Bckg. sub.) 
uncertainties refers to the 0--88\% $d+$Au spectra; the 
uncertainties are greater in the more peripheral $d+$Au and more central 
$N+$Au spectra. }
\begin{ruledtabular}
\begin{tabular}{cccccc}
$p_T$\,(GeV/$c$) & Mom.        & Mom.   & Bckg.      & $\delta_{\pi \rm{loss}}$ (\%)& Total(\%)\\ 
               & res. (\%) & scale (\%) & sub. (\%)&  &  \\ \hline
$<$4.5& $<$0.5& $<$3.2& $<$0.1& $<$0.3& $<$3.3\\
4.5--5.5& $<$0.6& 3.3& 0.5& 0.5& 3.4\\
5.5--6.5& 0.8& 3.5& 1.4& 1.1& 4.0\\
6.5--7.5& 1.0& 3.6& 2.0& 3.6& 5.6\\
7.5--8.5& 1.4& 3.7& 4.9& 6.9& 9.3\\
8.5--9.5& 1.8& 3.8& 11.9& 13.9& 18.8\\
\end{tabular}  
\end{ruledtabular}
\end{table}


We estimate systematic uncertainties in the methods and assumptions of 
our analysis as displayed in Tables~\ref{systerr} and~\ref{systerrpt}.  
Table~\ref{systerr} shows uncertainties in the spectra that do not vary 
with $p_{T}$, and Table~\ref{systerrpt} shows uncertainties that vary 
with $p_T$.  All the uncertainties listed in Table~\ref{systerr} 
propagate into the $R_{AB}$ and the hadron/$\pi^{0}$ ratio, although the 
uncertainty from the feed-down correction partially cancels in the 
$R_{AB}$ because of an analogous uncertainty in the $p+p$ reference spectrum.  
The background subtraction and momentum scale uncertainties only 
partially cancel in the $R_{AB}$; the momentum resolution uncertainty and 
$\delta_{\pi \rm{loss}}$ fully cancel.  The $p+p$ reference spectrum also 
introduces an uncertainty that ranges from 10.7\% in the low \pt\ bins to 
11.3\% at \pt\ $>$2 GeV/c.  The \pizero\ analysis~\cite{ppg044} we used in 
the hadron/\pizero\ ratios applies the same BBC bias correction factors as 
the present analysis and therefore the uncertainty stemming from this 
correction factor is canceled.

\section{Results\label{sec:results}}

The fully corrected \pt\ distributions of ($h^+ + h^-)/2$ for $d+$Au and 
$N+$Au collisions for minimum bias and four centrality classes are shown in 
Fig.~\ref{fig:data}.  The $d+$Au data are in good agreement with the sum of 
identified hadrons published in Ref. ~\cite{felix}.


Figure~\ref{fig:RAB} shows $R_{d\rm{Au}}$ and $R_{N\rm{Au}}$.  As expected, in the 
most peripheral $N+$Au bin, with $N_{\rm{coll}}=2.1$, $R_{N\rm{Au}}$ is consistent 
with unity.  $R_{d\rm{Au}}$ and $R_{N\rm{Au}}$ are in agreement within our 
uncertainty bounds.

The enhancement of the hadron yield relative to $p+p$ collisions has 
previously been observed in lower energy $p+$A 
collisions~\cite{lowenergy,cronin}, and is called the Cronin effect.  We 
observe that $R_{d\rm{Au}}$ and $R_{N\rm{Au}}$ are systematically larger than unity in 
the momentum range between 1.5\,\gevc\ and 5\,\gevc\ with maximum amplitude 
around 1.3.

There are many theoretical models with very different assumptions about 
initial state effects, which describe the Cronin effect~\cite{zhang, 
multiple, hard, accardi}.  All the models agree that there is at least 
one additional scattering of the initial nucleon or parton while 
propagating though the target nucleus. This scattering increases the 
intrinsic transverse momentum of the colliding parton, and leads to a 
broadening of the parton \pt\ distribution. We can parametrize the effect 
of this broadening by writing the mean value of parton intrinsic momentum 
$k_T$ as
\begin{equation}
\langle k^2_T\rangle_{pA} = \langle k^2_T\rangle_{pp} + \langle k^2_T\rangle_{A}, 
\label{eq:kT}
\end{equation}
where $\langle k^2_T\rangle_{pp}$ is the square of the initial parton 
transverse momentum in the proton, $\langle k^2_T\rangle_{A}$ is an 
additional momentum squared after rescattering, and $\langle 
k^2_T\rangle_{pA}$ is the final broadened width. Most of the models 
differ on the assumption they use to describe $\langle k^2_T\rangle_{A}$: 
whether there is a single hard scattering~\cite{hard} or a sum of small 
sequential rescatterings~\cite{multiple} that produces the additional 
$k_T$.

Common to the models is that $\langle k^2_T\rangle_{A}$ is a function of 
the number of sequential nucleon-nucleon collisions, $\nu$. For impact 
parameter $b$, $\langle k^2_T\rangle_{A}$ can be written as:
\begin{equation}
\langle k^2_T\rangle_{A}(b) = H (\nu(b) -1) , 
\label{eq:kT2}
\end{equation} 
where $H$ is the square of the average momentum acquired in $\nu$ -- 1
rescatterings. For a single hard scattering model, $\langle
k^2_T\rangle_{A}$ should saturate at $\nu$ = 2.  We therefore
investigate the shape of $R_{d\rm{Au}}$ as a function of $\nu$ to illuminate the
underlying process. The centrality selection of our data and the
tagged $N+$Au sample allow us to investigate precisely the effect of
the collision geometry.  We use $\nu = \langle N_{\rm{coll}} /
N^{\rm{deutron}}_{\rm{part}} \rangle$ (in $N+$Au collisions $\nu = \langle
N_{\rm{coll}}\rangle$) to look explicitly at the impact parameter
dependence of the nuclear modification factor.  

In Fig.~\ref{fig:nu}, we plot $R_{d\rm{Au}}$ and $R_{N\rm{Au}}$ as a function of 
$(\nu-1)$.  The values of $\nu$ are presented in Table~\ref{tablenn}.

Three transverse momentum regions were selected to study the dependence 
of $R_{AB}$ on $\nu$: 2.8$\leq$ \pt\ $\leq$ 6.0, 1.5$\leq$ 
\pt\ $\leq$ 2.7, and 0.6$\leq$ \pt\ $\leq$ 1.0 GeV/$c$.  In the low 
\pt\ region, we expect scaling with the number of participating nucleons rather 
than with the number of binary collisions; therefore $R_{AB}$ is less 
than unity. The Cronin effect is observed in the 2.8$\leq$ \pt\ $\leq$ 6.0 
\gevc\ region, where within the limits of our uncertainties, it is 
independent of the number of additional scatterings ($\nu$ -- 1).  In the 
intermediate \pt\ region, the data show little to no Cronin enhancement, thus 
confirming scaling with the number of binary collisions.  Just as 
$R_{d\rm{Au}}$($p_{\rm T}$) and $R_{N\rm{Au}}$($p_{\rm T}$) matched very closely, so 
do $R_{d\rm{Au}}(\nu)$ and $R_{N\rm{Au}}(\nu)$.

As discussed above, previous experiments found a larger Cronin 
enhancement for protons than pions ~\cite{lowenergy}.  Recently the 
PHENIX Collaboration published identified pion, kaon, and proton 
production data from $\sqrt{s_{NN}}=200$\,GeV $d+$Au 
collisions~\cite{felix}, in which a similar Cronin behavior was observed.  
$R_{d\rm{Au}}$ for protons reaches about 1.8 at $p_T \approx 3$\,GeV/$c$, 
whereas pion $R_{d\rm{Au}}$ is measured to be about 1.1 for transverse 
momentum between 2 and 2.6\,GeV/$c$.

At the time these data were taken, the PHENIX experiment did not have the 
capability to measure identified charged pions and protons at momenta 
above 2.6\,\gevc\ and 4\,GeV/$c$, respectively.  To extend the trends 
observed in these low momentum regions~\cite{felix}, we calculate the 
ratio of the charged hadrons measured in the present analysis to the 
$\pi^{0}$ spectra from Ref. ~\cite{ppg044}, alongside the ratio of 
$(p^{\pm}+K^{\pm}+\pi^{\pm})/\pi^{\pm}$ from~\cite{felix}.  These ratios 
are presented in Fig.~\ref{fig:pihad}.

In the transverse momentum region common to the two presented ratios 
there is strong agreement between the analyses.  The 
$(h^{+}+h^{-})/2\pi^{0}$ ratio is independent of \pt\ above 2\,\gevc\ where 
the identified particle data end, implying that the $R_{d\rm{Au}}$ and 
particle ratio trends observed at low transverse momentum continue at 
higher transverse momentum.  The average value of 1.58 $\pm$ 0.03 of the 
$(h^{+}+h^{-})/2\pi^{0}$ ratio for \pt\ above 2\,\gevc\ in the peripheral D 
centrality bin agrees well with the value of 1.59 obtained from lower 
energy collisions in the CERN Intersecting Storage Rings (ISR) ~\cite{ISR}.  As found in Ref. ~\cite{ppg003} 
for Au+Au collisions, this value rises for more central events: we find 
average values of 1.78 $\pm$ 0.02, 1.77$\pm$ 0.03, and 1.72 $\pm$ 0.03 
for centrality bins A, B, and C, respectively. There is an additional 
11\% systematic uncertainty common to all four values. The centrality 
dependence implies moderate medium modification effects in central 
$\sqrt{s_{NN}}=200$\,GeV collisions, even in the $d$+Au system, that 
increase the production of protons and kaons relative to pions. The 
PHENIX measurement of particle species dependent $R_{d\rm{Au}}$ 
in~\cite{felix} at lower \pt\ suggests that the increased particle 
production relative to pions is dominantly proton and not kaon 
production.

\section{Summary\label{sec:concl}}

We have measured the centrality dependence of charged hadron production 
at midrapidity in $d$+Au collisions, as well as in $p$+Au and $n$+Au 
collisions that are tagged by a spectator nucleon from the deuteron 
nucleus. The hadron yields in $p$+Au and $n$+Au collisions are identical 
within our experimental uncertainty. Using the $p+p$ data from same 
energy, we calculated the nuclear modification factors $R_{{dAu}}$ and 
$R_{{NAu}}$ for various centrality selections.  Within this analysis and 
its experimental uncertainty, there is no difference between $R_{{dAu}}$ 
and $R_{{NAu}}$ . Instead of a strong suppression as predicted by some 
color glass condensate models~\cite{colorglass}, an excess of hadron 
production is seen at \pt\ $>$2 GeV/$c$, consistent with enhancement due to 
the Cronin effect. The magnitude of the Cronin effect is independent of 
the number of additional scatterings ($\nu$ -- 1) within the limits of our 
uncertainties.  We also studied the ratio of charged hadron yield to pion 
yield. We found that the charged pions account for about 60\% of the 
charged hadrons at \pt\ $>$3 GeV/$c$, with a slightly larger value in 
central $d$+Au collisions. This implies that $R_{{dAu}}$ for protons and 
kaons remains close to one at higher $p_{T}$.


\begin{acknowledgments}


We thank the staff of the Collider-Accelerator and Physics
Departments at Brookhaven National Laboratory and the staff of
the other PHENIX participating institutions for their vital
contributions.  We acknowledge support from the 
Office of Nuclear Physics in the
Office of Science of the Department of Energy, the
National Science Foundation, Abilene Christian University
Research Council, Research Foundation of SUNY, and Dean of the
College of Arts and Sciences, Vanderbilt University (U.S.A);
Ministry of Education, Culture, Sports, Science, and Technology
and the Japan Society for the Promotion of Science (Japan);
Conselho Nacional de Desenvolvimento Cient\'{\i}fico e
Tecnol{\'o}gico and Funda\c c{~a}o de Amparo {\`a} Pesquisa do
Estado de S$\tilde{\rm{a}}$o Paulo (Brazil);
Natural Science Foundation of China (People's Republic of China);
Centre National de la Recherche Scientifique, Commissariat
{\`a} l'{\'E}nergie Atomique, and Institut National de Physique
Nucl{\'e}aire et de Physique des Particules (France);
Ministry of Industry, Science and Tekhnologies,
Bundesministerium f\"ur Bildung und Forschung, Deutscher
Akademischer Austausch Dienst, and Alexander von Humboldt Stiftung (Germany);
Hungarian National Science Fund, OTKA (Hungary), 
Department of Atomic Energy (India); 
Israel Science Foundation (Israel); 
Korea Research Foundation 
and Korea Science and Engineering Foundation (Korea);
Ministry of Education and Science, Russia Academy of Sciences,
Federal Agency of Atomic Energy (Russia),
VR and the Wallenberg Foundation (Sweden); 
the U.S. Civilian Research and Development Foundation for the
Independent States of the Former Soviet Union; the US-Hungarian
NSF-OTKA-MTA; and the U.S.-Israel Binational Science Foundation.

\end{acknowledgments}



\def\NIMA{Nucl. Instrum. Meth. A}


\begin{thebibliography}{99}

\bibitem{ppg003} K.~Adcox {\it et al.} (PHENIX Collaboration),  
Phys.\ Rev.\ Lett. {\bf 88}, 022301 (2002).

\bibitem{ppg014} S.~S.~Adler {\it et al.} (PHENIX Collaboration),  
Phys.\ Rev.\ Lett. {\bf 91}, 072301 (2003).

\bibitem{ppg023} S.~S.~Adler {\it et al.} (PHENIX Collaboration),  
Phys.\ Rev.\ C {\bf 69}, 034910 (2004).

\bibitem{en_loss} M.~Gyulassy and M.~Plumer, 
Phys.\ Lett.\ {\bf B243}, 432 (1990); 
X.~N.~Wang and M.~Gyulassy, 
Phys.\ Rev.\ Lett. {\bf 68}, 1480 (1992).

\bibitem{jet_modif} S.~S.~Adler {\it et al.} (PHENIX Collaboration), 
Phys.\ Rev.\ Lett. {\bf 97}, 052301 (2006). 

\bibitem{ppg033} S.~S.~Adler {\it et al.} (PHENIX Collaboration), 
Phys.\ Rev.\ C {\bf 71}, 051902 (2005). 

\bibitem{Adler:2003kg} S.~S.~Adler {\it et al.} (PHENIX Collaboration), 
Phys.\ Rev.\ Lett.\  {\bf 91}, 172301 (2003)

\bibitem{recomb} V.~Greco, C.~M.~Ko and P.~L$\acute{e}$vai, 
Phys.\ Rev.\ Lett.\ {\bf 90}, 202302 (2003).

\bibitem{colorglass} D.~Kharzeev, Yu.~V.~Kovchegov and K.~Tuchin, 
Phys.\ Rev.\ D {\bf 68}, 094013 (2003).

\bibitem{cronin} J.~W.~Cronin  {\it et al.}, 
Phys.\ Rev.\ D {\bf 11}, 3105 (1975).

\bibitem{lowenergy} P.~B.~Straub  {\it et al.}, 
Phys.\ Rev.\ Lett. {\bf 68}, 452 (1992).

\bibitem{minbias} S~.S.~Adler  {\it et al.} (PHENIX Collaboration), 
Phys.\ Rev.\ Lett.\ {\bf 91}, 072303 (2003).

\bibitem{ppg044} S~.S.~Adler  {\it et al.} (PHENIX Collaboration), 
Phys.\ Rev.\ Lett.\ {\bf 98}, 172302 (2007).

\bibitem{lev} M.~Lev and B.~Peterson, Z.\ 
Phys. C \ {\bf 21}, 155 (1983).

\bibitem{NIM} K.~Adcox  {\it et al.} (PHENIX Collaboration),  
\NIMA {\bf 499}, 469-602 (2003).

\bibitem{MC}  S~.S.~Adler  {\it et al.} (PHENIX Collaboration),  
\NIMA {\bf 499}, 593 (2003).

\bibitem{zdc} C.~Adler  {\it et al.} (STAR Collaboration), 
\NIMA {\bf 499}, 433 (2003).

\bibitem{e864} T.~A.~Armstrong  {\it et al.}, 
\NIMA {\bf 406}, 227 (1998).

\bibitem{white} S.~White, 
AIP Conf.Proc. {\bf 792}, 527 (2005).

\bibitem{nagle}  S~.S.~Adler {\it et al.} (PHENIX Collaboration), 
Phys.\ Rev.\ Lett.\ {\bf 94}, 082302 (2005).

\bibitem{glaubparam} L.~Hulth$\acute{\rm{e}}$n and M.~Sagawara, Handbuch\ 
der\ Physik, edited by S. Fl$\ddot{\rm{u}}$gge (Springer-Verlag, New York, 1957), Vol. 39.\ {\bf 39}, (1957).

\bibitem{new_param} A.~F.~Krutov and V.~E.~Troitsky, 
Phys.\ Rev.\ C {\bf 76}, 017001 (2007).

\bibitem{glauber} R.~J.~Glauber and G.~Matthiae, 
Nucl.\ Phys.\ {\bf B21}, 135 (1970).

\bibitem{pythia} T.~Sj$\ddot{o}$strand, L.~L$\ddot{o}$nnbland and S.~Mrenna, 
arXiv:hep-ph/0108264 (2001).

\bibitem{an421} V.~Ryabov [for PHENIX Collaboration], 
Nucl.\ Phys.\ {\bf A774}, 735 (2006).

\bibitem{felix} S~.S.~Adler  {\it et al.} (PHENIX Collaboration), 
Phys.\ Rev.\ C {\bf 74}, 024904 (2006).

\bibitem{star} J.~Adams  {\it et al.} [STAR Collaboration], 
Phys.\ Lett.\ {\bf B637}, 161 (2006).

\bibitem{ppg26}S~.S.~Adler  {\it et al.} (PHENIX Collaboration), 
Phys.\ Rev.\ C {\bf 69}, 034909 (2004).

\bibitem{ppref}S.~S.~Adler  {\it et al.} (PHENIX Collaboration), 
Phys.\ Rev.\ Lett.\ {\bf 95}, 202001 (2005)

\bibitem{zhang} Y.~Zhang  {\it et al.}, 
Phys.\ Rev.\ C {\bf 65}, 034903 (2002).

\bibitem{multiple} X.~N.~Wang, 
Phys.\ Rev.\ C {\bf 61}, 064910 (2000);
Y.~Zhang  {\it et al.}, 
Phys.\ Rev.\ C {\bf 65}, 034903 (2002).

\bibitem{hard} B.~Z~Kopeliovich  {\it et al.}, 
Phys.\ Rev.\ Lett.\ {\bf 88}, 232303 (2002).
A.~Accardi and D.~Treleani, 
Phys.\ Rev.\ D {\bf 64}, 116004 (2001). I.~Vitev and M.~Gyulassy, 
Phys.\ Rev.\ Lett.\ {\bf 89}, 252301 (2002)

\bibitem{accardi} A.~Accardi, arXiv: hep-ph/0212148.


\bibitem{ISR} B.~Alper {\it et al.}  (British-Scandinavian Collaboration), 
Nucl.\ Phys.\ {\bf B100}, 237 (1975).

\end{thebibliography}
\end{document}